\documentclass{aastex63}

\newcommand{\msun}{$\rm M_\odot$}

\newcommand\nuk[2]{$\rm ^{\rm #2} #1$}

\received{.............}
\revised{.............}
\accepted{.............}
\submitjournal{ApJ}
\shorttitle{Core mass-radius relation of massive stars}
\shortauthors{Chieffi,Limongi}
\begin{document}
\title{The presupernova core mass-radius relation of massive star: understanding its formation and evolution.}
\author[0000-0002-3589-3203]{Alessandro Chieffi}
\affiliation{Istituto Nazionale di Astrofisica - Istituto di Astrofisica e Planetologia Spaziali, Via Fosso del Cavaliere 100, I-00133, Roma, Italy; alessandro.chieffi@inaf.it}
\affiliation{Monash Centre for Astrophysics (MoCA),
School of Mathematical Sciences, Monash University, Victoria 3800, Australia}
\author[0000-0003-0636-7834]{Marco Limongi}
\affiliation{Istituto Nazionale di Astrofisica - Osservatorio Astronomico di Roma, Via Frascati 33, I-00040, Monteporzio Catone, Italy; marco.limongi@inaf.it}
\affiliation{Kavli Institute for the Physics and Mathematics of the Universe, Todai Institutes for Advanced Study, the University of Tokyo, Kashiwa, Japan 277-8583 (Kavli IPMU, WPI)}

\correspondingauthor{Alessandro Chieffi}
\email{alessandro.chieffi@inaf.it}

\begin{abstract}
We present a fine grid of solar metallicity models of massive stars (320 in the range 12$\leq$M(\msun)$\leq$27.95), extending from the Main Sequence up to the onset of the collapse, in order to quantitatively determine how their compactness $\xi_{2.5}$ (as defined by O'Connor $\&$ Ott, 2011, ApJ 730, 70) scales with the Carbon Oxygen core mass at the beginning of the core collapse. We find a well defined, not monotonic (but not scattered) trend of the compactness with the Carbon Oxygen core mass that is strictly (and mainly) correlated to the behavior, i.e. birth, growth and disappearance, of the various C convective episodes that follow one another during the advanced evolutionary phases. Though both the mass size of the Carbon Oxygen core and the amount of \nuk{C}{12} left by the He burning play a major role in sculpting the final Mass-Radius relation, it is the abundance of \nuk{C}{12} the ultimate responsible for the final degree of compactness of a star because it controls the ability of the C burning shell to advance in mass before the final collapse. 

\end{abstract}

\section{Introduction} \label{sec:intro}
A proper understanding of the final fate of a massive star is mandatory to estimate some of the outcomes of its explosion like, e.g., the mass of the remnant and the chemical composition of the ejecta. In order to reach such a goal, both the presupernova evolution and the following explosion must be properly simulated.

In the last decade, the largest body of theoretical works devoted to the explosion of a massive star was mainly focused on  the progressively more sophisticated treatment of the neutrino transport in multidimensional hydrodynamic simulations of the core collapse. Given the enormous amount of literature on the subject we refer the reader to the leading groups that currently explore the explosion of a massive star in 3D \citep{bu19,ja17,mu17} and references therein.

On the other side, also the presupernova evolution is crucial because it determines some of the properties of the star at the onset of the core collapse that drive the following explosion like, e.g., the density profile [or, equivalently, the Mass-Radius (M-R) relation], the mass of the iron core and its electron fraction ($\rm Y_e$) profile \citep{coo84,ba85,be90,oo11,oo13}. 
Such a final configuration is the result of the complex interplay among the various nuclear burning and the number, timing and overlap of the various convective zones. In this context, one of the key uncertainties connected with this complex behavior is the treatment of the various instabilities (thermal, rotation induced, etc.) that, in most cases, are still simulated very crudely by means of the Schwarzchild/Ledoux criterion, the mixing length theory, presence/absence of convective overshooting, parametrized efficiency of semiconvection and so on. Given the large variety of different possible choices it is clear that the final structure of a star may depend, even significantly, on the choices adopted by each author/group. Moreover most of the computations presently available usually present results with a step in mass of at least half a solar mass or more (our typical step is of the order of a few solar masses).
However, in these last years the situation changed substantially because \cite{sw14} and \cite{su18} started a detailed study of the evolution of the massive stars and associated explosions by adopting a very fine step in mass ($\Delta$M=0.01\msun). Among the various results presented in these papers, an interesting outcome highlighted by the authors is that even minor changes in the initial mass of a star may lead to very different structures at the beginning of the collapse. Such strong variations in the density profile are readily visible by taking advantage of a parameter, firstly introduced by \cite{oo11}, that summarizes the compactness of a star by means of a single parameter $\xi$, that is just the ratio between the mass M and its corresponding radius R at the mass location M=2.5\msun, i.e. $\xi_{2.5}$=2.5\msun$\rm / R_{2.5}$(1000~km). Figure 8 in \cite{su18} shows exactly such a result. In particular between 14 and 20\msun~ and between 22 and 24\msun~ a large scatter in the compactness of the models is evident.

Since our first paper on the subject \citep{cls98} we have addressed many aspects of the evolution of the massive stars in a wide mass range (typically in the range 11 to 120\msun) and metallicity (0 to solar) \citep{lc12} and also various initial rotation velocities \citep{lc18}. Our typical step in mass has always been of the order of 1 solar mass  or more. Given the relevant implications of the results obtained by Sukhbold and coauthors, we consider of great interest compute, show and discuss the trend we do obtain for the $\xi$ parameter as a function of the initial mass with a mass step much smaller than used in our previous computations.

We will not attempt any connection between compactness and explodability because it is both beyond the purposes of the present study and also because it has often been criticized.  \cite{er16}, for example, proposed the adoption of two parameters to infer the possible explodability of a stellar model: the mass location and its derivative with respect to the radius evaluated at the coordinate where the entropy per nucleon reaches a value of 4 (which basically corresponds to the base of the O burning shell). We refer the reader to that paper for more details. Also \cite{bu19} regard as non reliable the use of the compactness $\xi$ to infer the explodability of a model. 

The paper is organized as follows. The version of the code adopted for this analysis is presented in Section \ref{sec:models} while the properties of all our models are discussed in details in Section \ref{sec:disc}. Section \ref{sec:comp} is devoted to a comparison between some of our results and those presented by \cite{su18}. A final conclusion summarizes our results. 

\section{The Models} \label{sec:models}

All the models discussed in this paper have been computed with the FRANEC evolutionary code, release 6. This version is the same used in \cite{lc18}, with the exception of the nuclear network and the number of mesh points. In this set of computations we adopted a reduced network (shown in table \ref{tab:net}) because we were basically interested in the physical evolution of the models and not in the detailed nucleosynthesis but also because the calculation of this very large grid of models with our full network would have required an unfeasible amount of computer time. However, in order to check the consequences of this choice, we computed four models with the full network and found that the final compactness $\xi$ (the main property we are interested in this contest) closely resembles the one obtained with the small network (see Section \ref{sec:comp}). The number of mesh points has been mildly increased so that they now range between 2000 and 6000 (apart from the outermost 1\% of the mass, i.e. the envelope, that is described by a few hundreds mesh points), depending on the mass and the evolutionary phase. A great effort was devoted in choosing a mesh distribution refined enough to provide a very clean temporal evolution of the central He burning, in order to avoid the spurious ingestion of fresh He towards the end of the He burning and hence a {\it random} scatter in the final C abundance. Figure \ref{fig:mcoc12} shows in the left panel the run of the central C abundance left by the core He burning as a function of the initial mass (red dots). A scatter, even modest, in the C abundance would spoil all the following advanced burning (because of the tremendous importance of the C abundance in driving all the advanced evolutionary phases) and therefore it would vanish all the efforts to produce a clean starting point for the advanced burning. Our grid of models consists of 320 evolutionary tracks in the range 12 to 27.95\msun~ with a step in mass of 0.05\msun. We adopted the solar metallicity of \cite{agss09} ($\rm Z=1.345\times 10^{-2}$), a He abundance equal to Y=0.265 and a mixing length parameter $\alpha=2.1$. Table \ref{tab:maindata} shows some relevant data of the models presented here. Columns 1 to 5 show, respectively, the initial mass, and the final values of the total mass, the He core mass, the CO core mass and the Fe core mass, all in solar masses. The last two columns show the final compactness $\xi$ evaluated for the CO core mass and 2.5\msun.  All models were evolved up to a central temperature of $\sim8$ Gk.

\begin{deluxetable}{cc}
\tablewidth{3 in}
\tablecolumns{2}
\tablecaption{Network \label{tab:net}}
\tablehead{isotope&isotope}
\startdata
\nuk{H}{1}   &  \nuk{Mg}{24}\\
\nuk{H}{2}   &  \nuk{Si}{28}\\
\nuk{He}{3}  &  \nuk{Na}{23}\\
\nuk{He}{4}  &  \nuk{P}{31} \\
\nuk{Li}{7}  &  \nuk{S}{32} \\
\nuk{Be}{7}  &  \nuk{Ar}{36}\\
\nuk{C}{12}  &  \nuk{Ca}{40}\\
\nuk{C}{13}  &  \nuk{Ca}{44}\\
\nuk{N}{13}  &  \nuk{Ti}{44}\\
\nuk{N}{14}  &  \nuk{Ti}{48}\\
\nuk{N}{15}  &  \nuk{Cr}{48}\\
\nuk{O}{15}  &  \nuk{Cr}{52}\\
\nuk{O}{16}  &  \nuk{Fe}{52}\\
\nuk{O}{17}  &  \nuk{Ni}{56}\\
\nuk{F}{17}  &  \nuk{Fe}{56}\\
\nuk{Ne}{20} &              \\
\enddata
\end{deluxetable}

\section{Discussion} \label{sec:disc}
\begin{figure*}
\epsscale{1.}
\plottwo{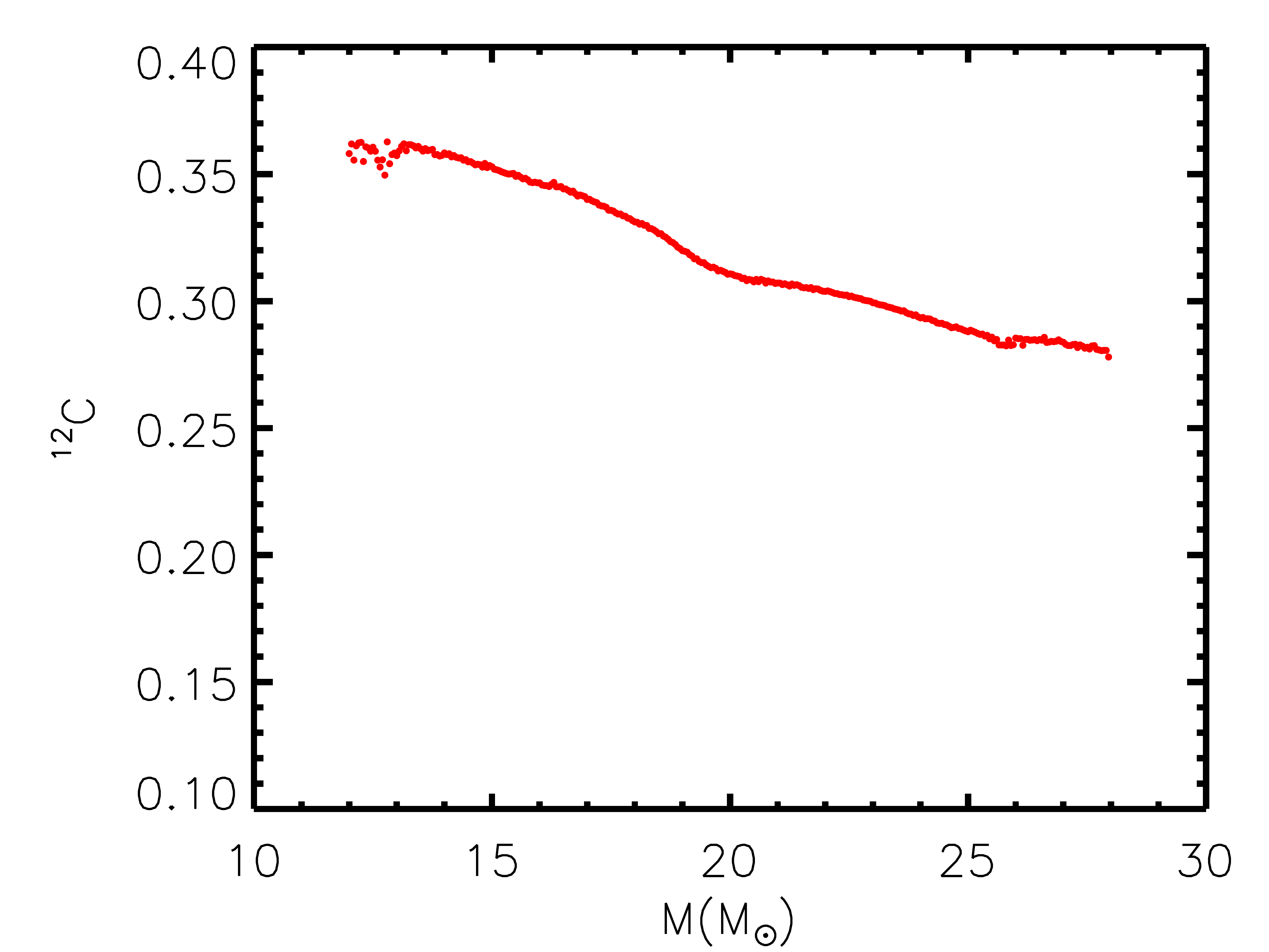}{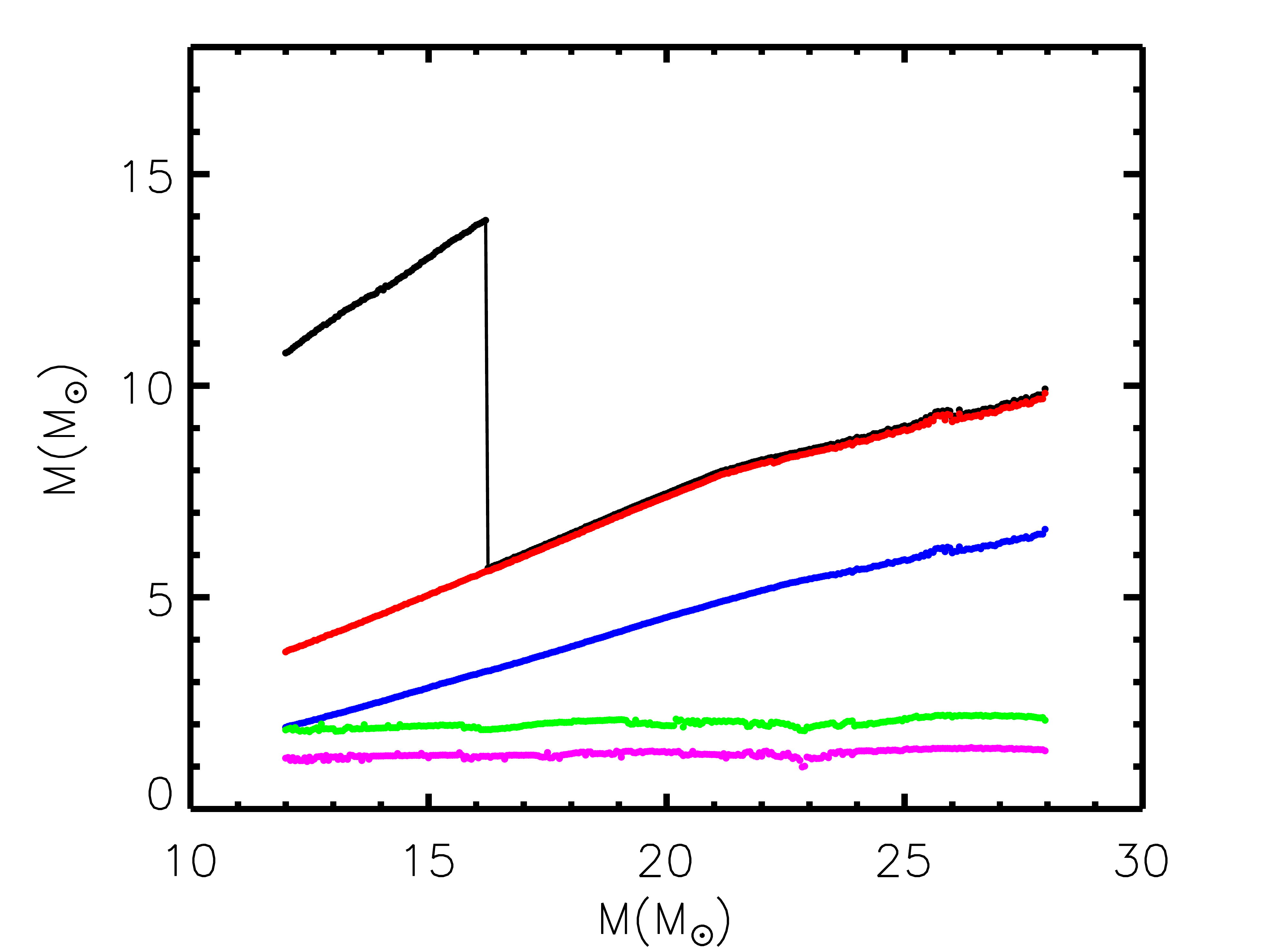}
\caption{(left) C mass fraction left by the He burning; (right) total mass (black), He core mass (red), CO core mass (blue), O burning shell (green) and Si burning shell (magenta). The O and Si burning shells have been shifted by +0.03\msun~ and by -0.03\msun, respectively, to improve readability. \label{fig:mcoc12}}
\end{figure*}

The advanced burning phases of a massive star, i.e. those going from the central He exhaustion up to the onset of the core collapse, are determined once both the CO core mass $\rm (M_{\rm CO})$ and the mass fraction of \nuk{C}{12} left by the central He burning are known. This means that, while the H and the He burning phases may be considered as mono-parametric, in the sense that they are controlled by a single parameter (the current mass or the He core mass), the advanced burning require the simultaneous knowledge of $\rm M_{\rm CO}$ and X(\nuk{C}{12}) in order to be uniquely determined and therefore may be considered bi-parametric. The CO core mass is fundamental because it plays the role the total mass has in central H burning and the He core mass has in He burning, while the \nuk{C}{12} left by the He burning determines the amount of fuel available to the C burning and hence determines the number and the extension (in mass) of the various C convective episodes: both contribute to shape the M-R relation at the onset of the core collapse and hence control the development of all the other burning and of the Fe core mass. Figure \ref{fig:mcoc12} shows in the right panel the run of the CO core mass (blue dots) as a function of the initial mass for all our 320 model. For sake of completeness, the same panel shows also the run of the total mass (black dots plus line),  the He core mass (red dots), the O burning shell (green dots) and the Si burning shell (magenta dots) with the initial mass. The vertical drop in the total mass occurring at M=16.25\msun~ marks the transition between stars that explode as red supergiants and those that turn blue before the final explosion. All four relations show a very tight dependence on the initial mass without basically any scatter.

\begin{figure}[ht!]
\epsscale{1.}
\plotone{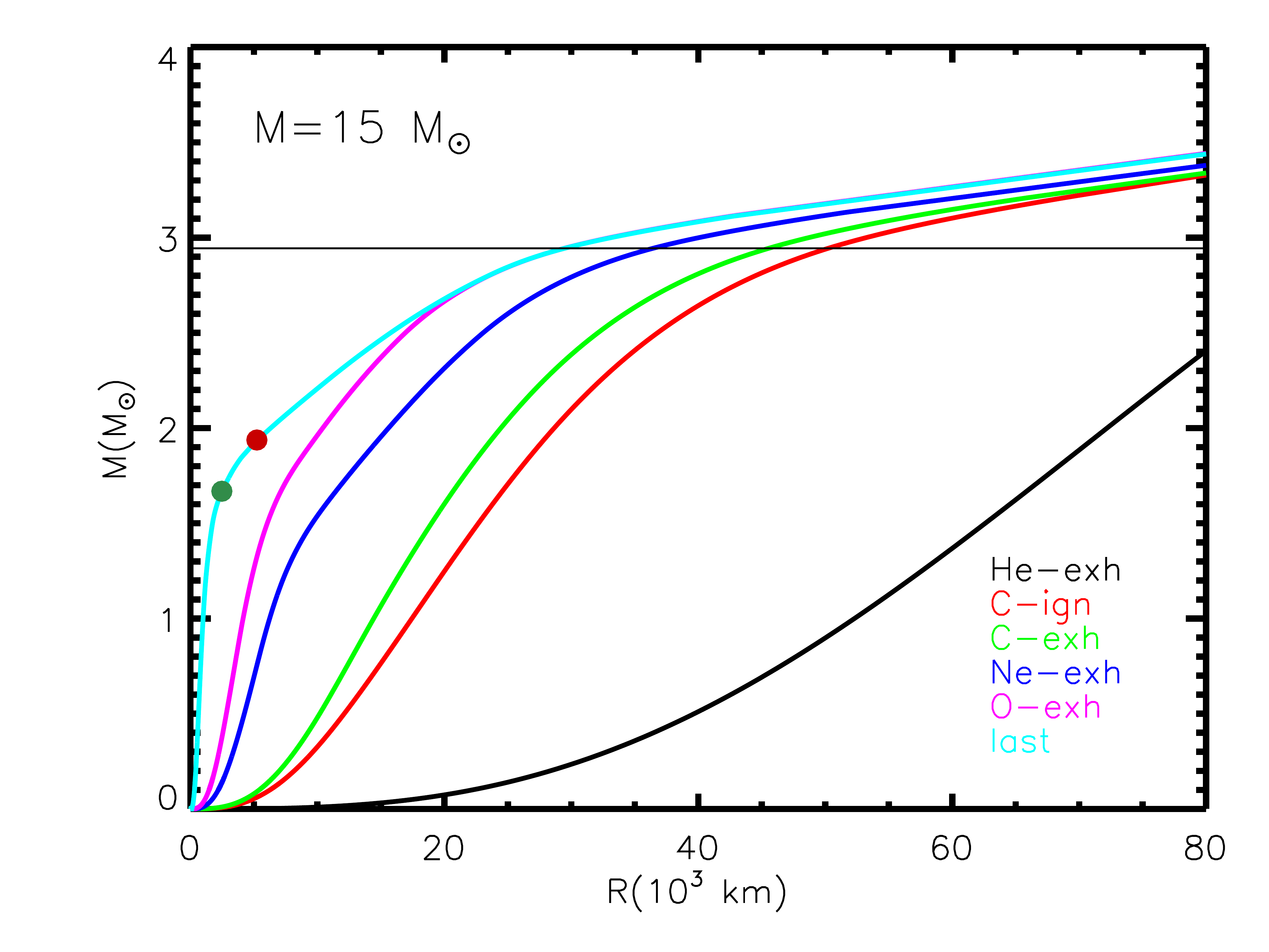}
\caption{Mass-Radius relation of a 15\msun~ of solar metallicity at different phases: He ignition (black), C ignition (red), C exhaustion (green), Ne exhaustion (blue), O exhaustion (magenta) and last model (cyan). The dark green and dark red dots mark, respectively, the bases of the O and the C burning shells in the last model. The thin black horizontal line shows the CO mass size.\label{fig:m15mr}}
\end{figure}

\begin{figure}[ht!]
\epsscale{1.}
\plotone{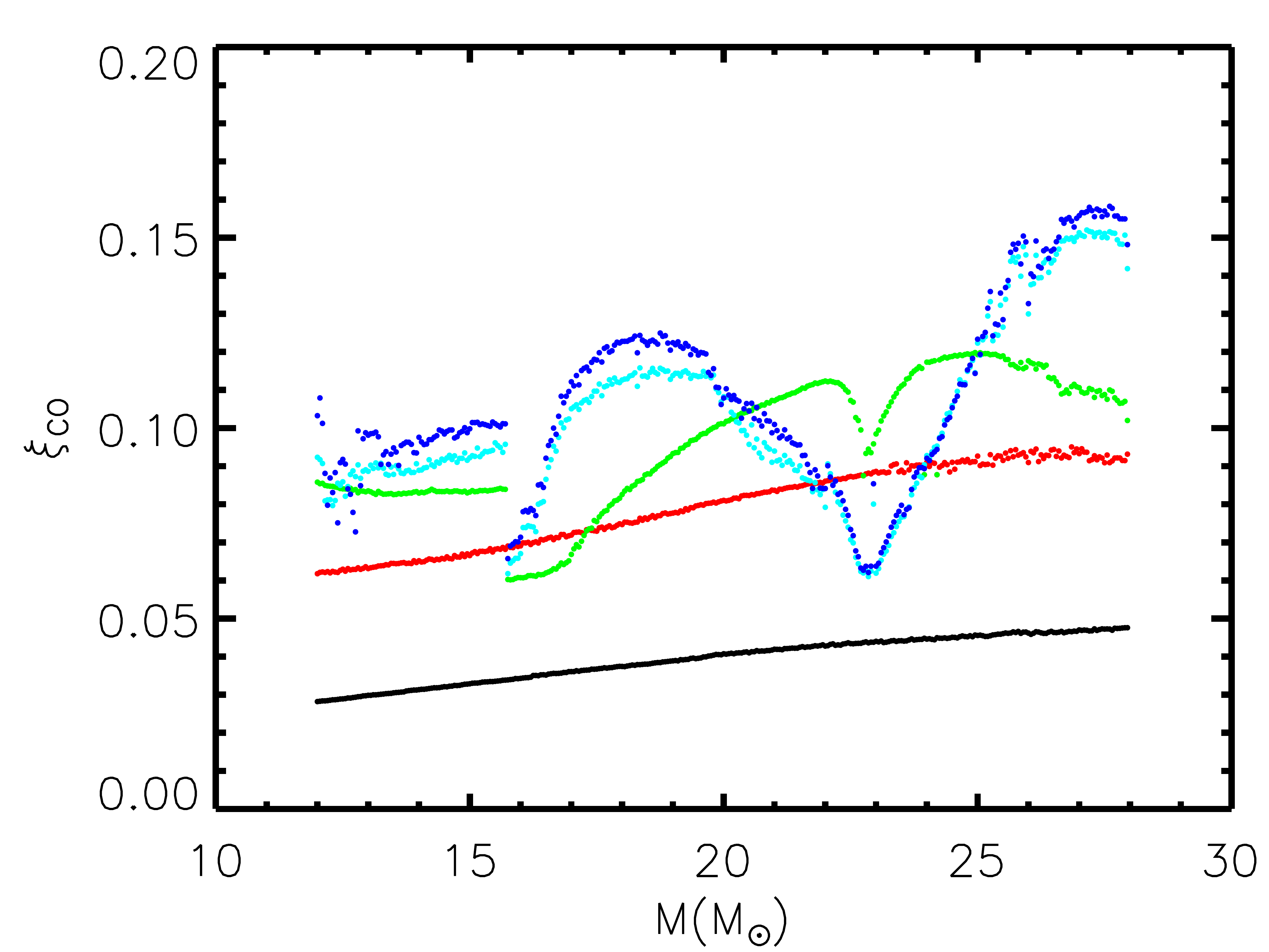}
\caption{Compactness of the CO core mass at various evolutionary phases: central He exhaustion (black),
central C exhaustion (red), central Ne ignition (green), central Si exhaustion (cyan) and last model (blue). \label{fig:csico}}
\end{figure}

In order to understand the scaling of the compactness of the stars with the initial mass at the onset of the core collapse, it is firstly necessary to fix an {\it operational} definition of the compactness of a star and then understand how it changes during its evolution. The natural relation that fully describes the compactness of a star in any evolutionary phase is the Mass-Radius (M-R) relation (or, equivalently, the density profile). Figure \ref{fig:m15mr} shows, as an example, the M-R relation of a 15\msun~ at various key evolutionary phases: the black line refers to the end of the central He burning, while the red, green, blue, magenta and cyan lines mark, respectively, the beginning and the end of the central C burning, the end of the central Ne and O burning and the last model. The dark green dot marks the position of the O burning shell (that practically coincides with the location where the entropy per barion S is equal to 4 in units of Boltzmann constant) while the dark red dot marks the position of the C burning shell. The black horizontal line marks the mass coordinate of the CO core. The smooth shallow M-R profile left by the He burning progressively steepens and a knee begins to appear as soon as an efficient burning shell forms. The main burning shell that controls the position and the bending of the knee just before the collapse is the O burning shell, how it is readily visible in Figure \ref{fig:m15mr}. This Figure may be considered a template since the M-R relation of any massive star shows a similar shape at the onset of the core collapse. Though this relation fully describes the compactness of a star, it is clear that it is not possible to compare the final M-R relations of all our 320 models in a single plot to determine its scaling with the initial mass. Therefore we decided to compare the compactness of some selected layers. In analogy with the strategy adopted by, e.g., \cite{oo11} we chose to define the compactness of any mass coordinate "$\rm M_i$" by means of the {\it operational} ratio $\rm \xi_{(i)}=M_i(M_\odot)/R_i(1000~km)$. 
The first relevant mass location worth being analyzed is the one corresponding to the the CO core, for which the compactness is defined as $\rm \xi_{CO}=M_{CO}(M_\odot)/R_{CO}(1000~km)$. The black dots in Figure \ref{fig:csico} show the run of $\rm \xi_{CO}$ with the initial mass soon after the formation of the CO core. At this stage a tight monotonic relation between the compactness of the CO core and the initial mass exists. The moderate increase of the M/R ratio with M is what one would qualitatively expect on the basis of dimensional arguments. In fact, a gas cloud in hydrostatic equilibrium has an M/R roughly constant if the equation of state (EOS) is dominated by the particles, while it scales as $\rm M^{1/2}$ if the EOS is dominated by photons. In a mixed case in which both particles and photons contribute significantly to the EOS, we expect a direct scaling of M/R with M. Full integration of the stellar equations confirms this qualitative expectation. This trend is not qualitatively modified by the central C burning, the only difference being an increase of the overall compactness of the CO core mass as a consequence of the natural continuous contraction of the core. So, at the end of the central C burning the scaling of $\rm \xi_{CO}$ with the mass is still tight and (almost) monotonic (red dots in Figure \ref{fig:csico}).
The (almost) monotonic relation between $\rm \xi_{CO}$ and initial mass disappears in the passage from the end of the central C burning to the beginning of the central Ne burning (green dots in Figure \ref{fig:csico}).  Though the correlation between the compactness of $\rm \xi_{CO}$ and the initial mass is still very tight, some {\it features} begin to appear. On average $\rm \xi_{CO}$ still increases with the initial mass, but now a jump forms at $\rm M_{\rm ini}$=15.75\msun, a minimum is present at $\rm M_{\rm ini}$=22.8\msun~ and a turn over occurs above 25\msun. The formation of these {\it features} reflects the different evolution of the C convective shells as the initial mass increases.

\begin{figure*}[ht!]
\gridline{\fig{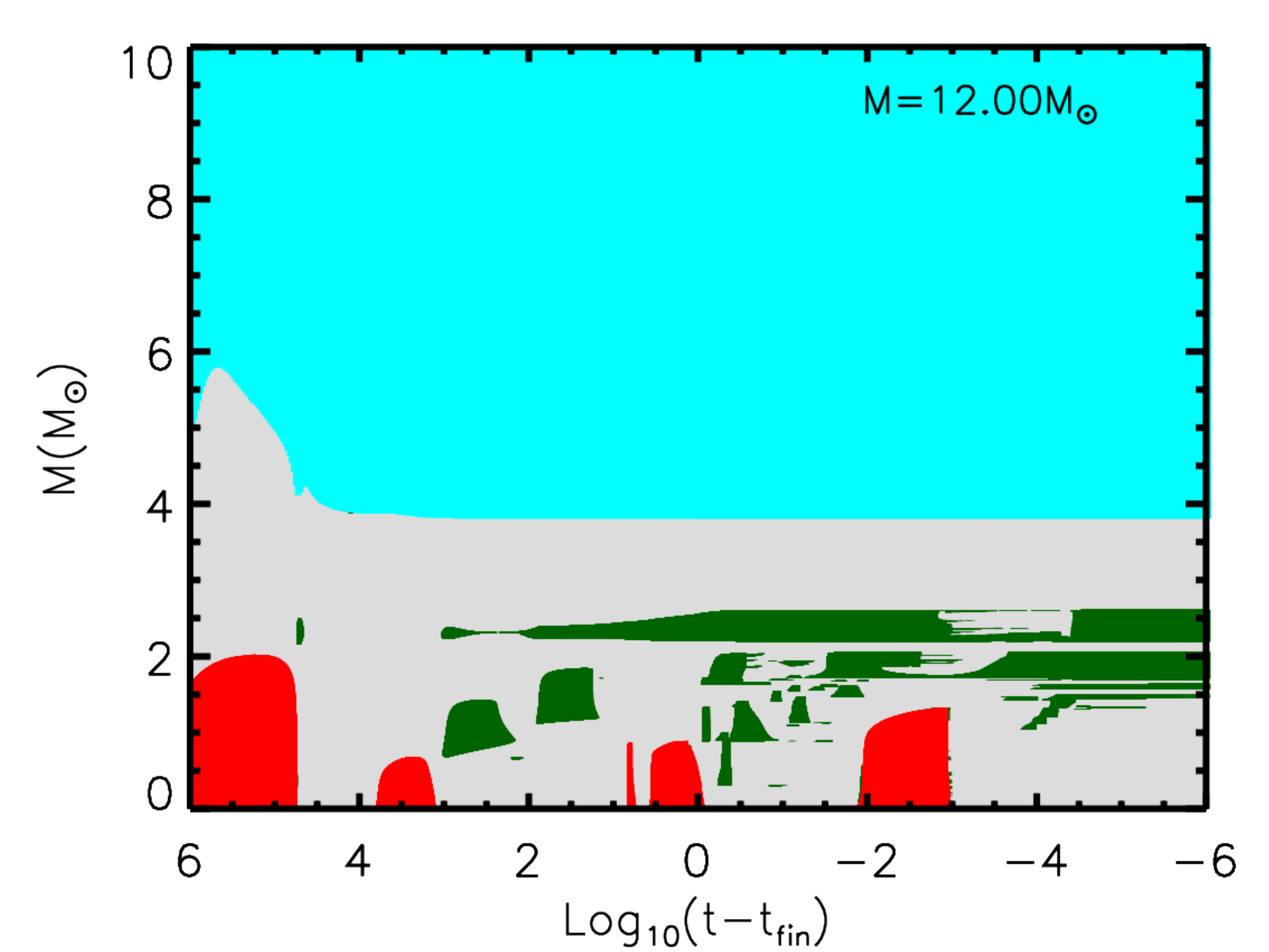}{0.30\textwidth}{(a)}
          \fig{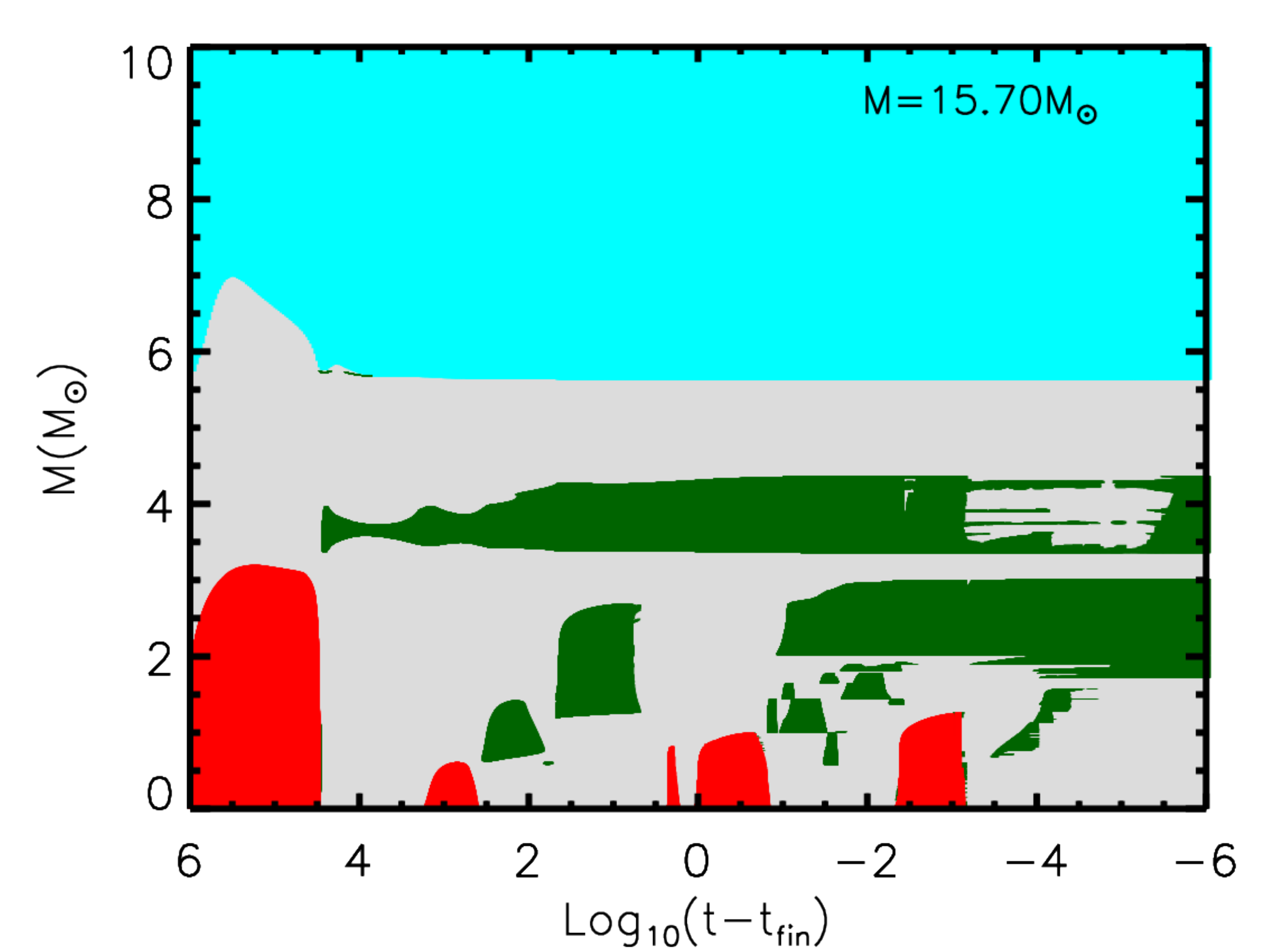}{0.30\textwidth}{(b)}
          \fig{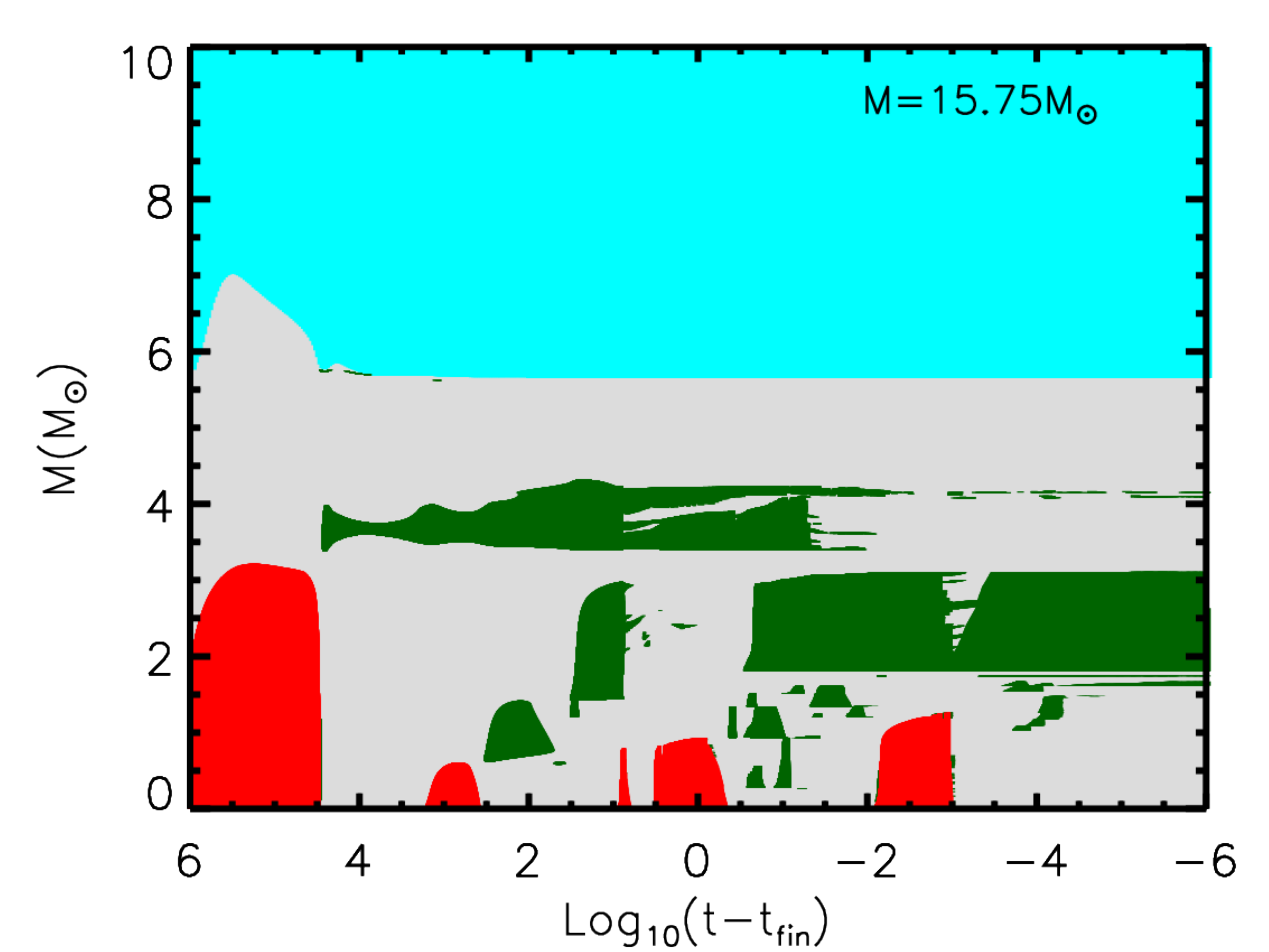}{0.30\textwidth}{(c)}}
\gridline{\fig{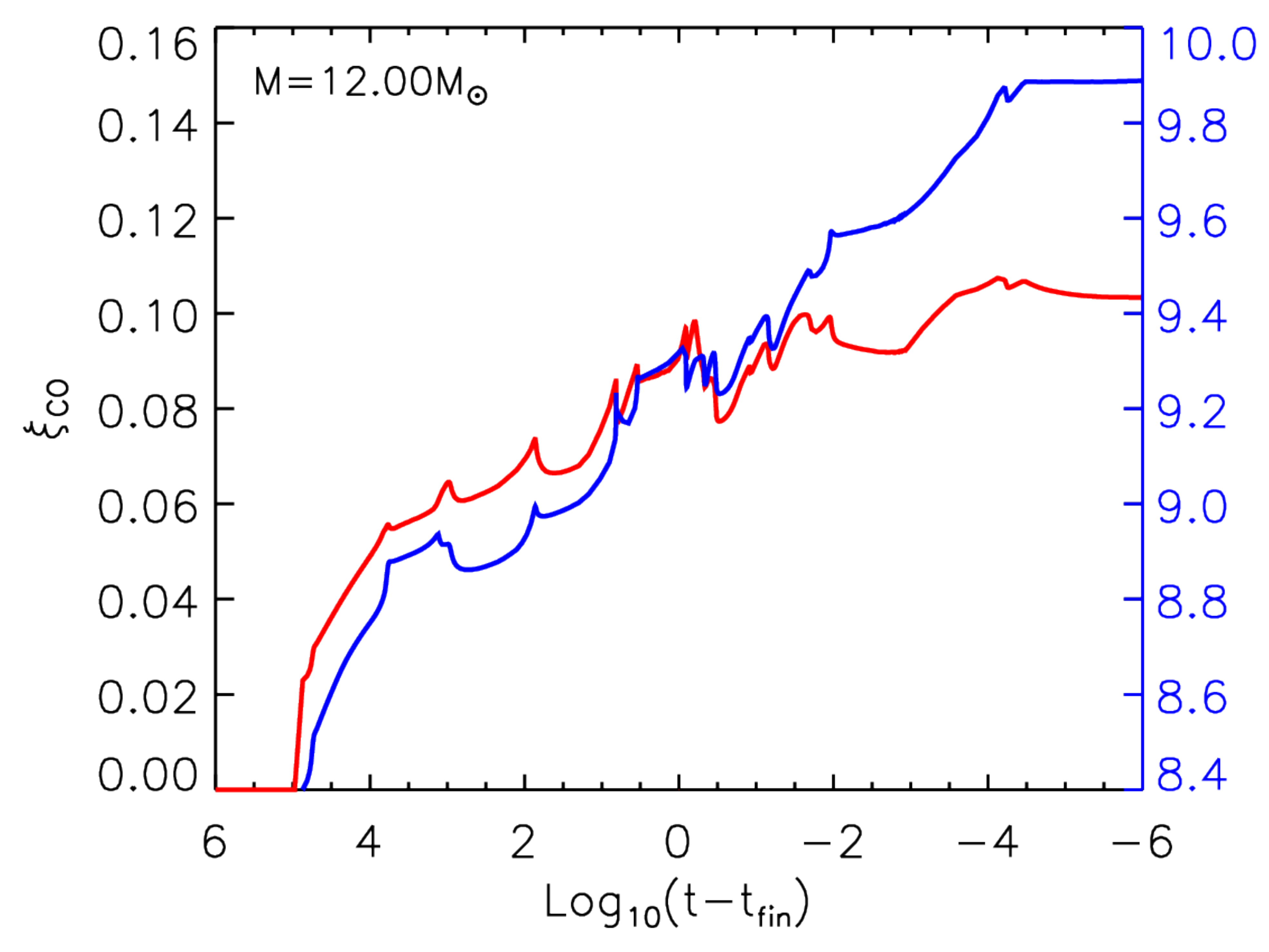}{0.33\textwidth}{(d)}
          \fig{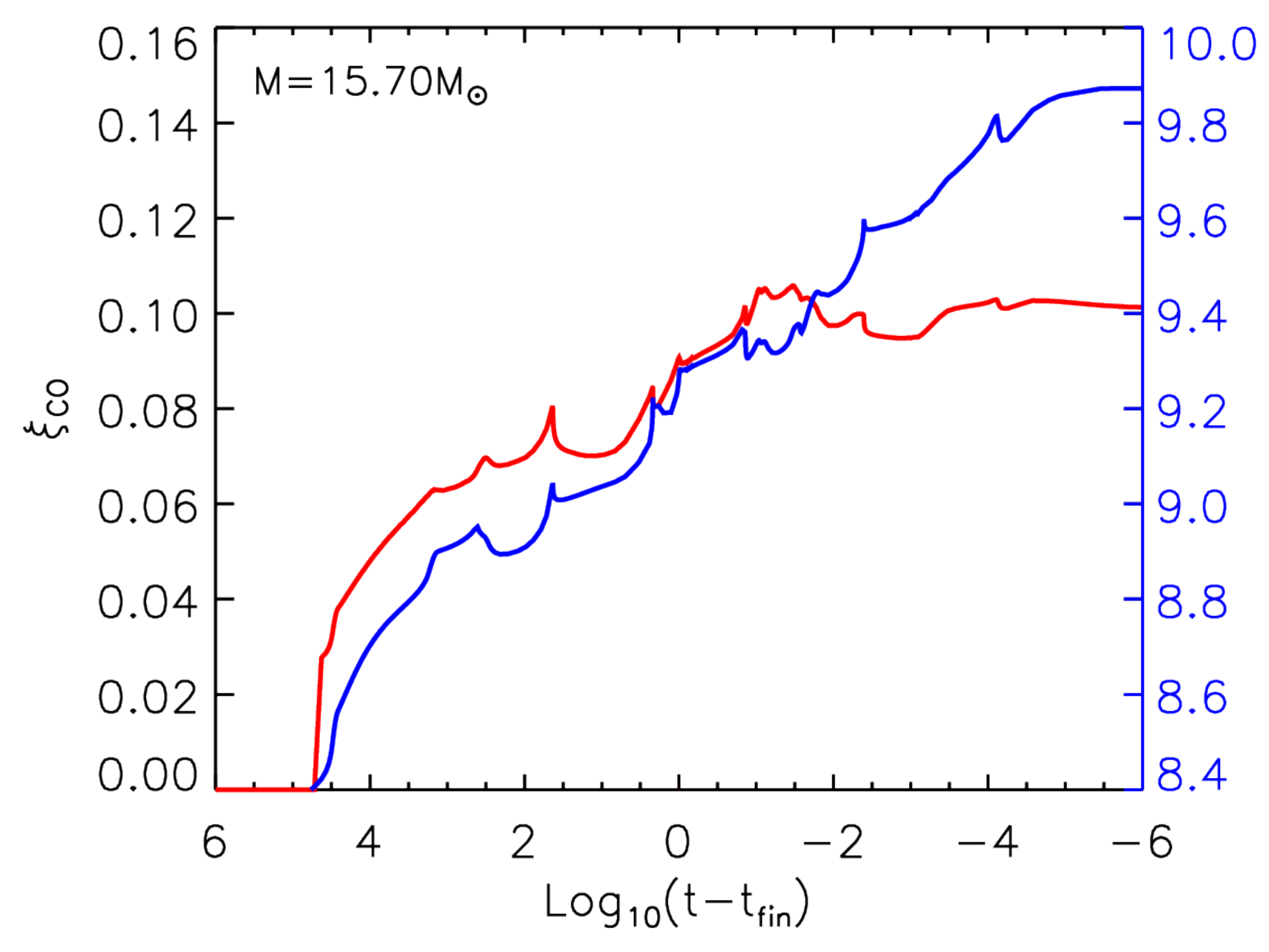}{0.33\textwidth}{(e)}
          \fig{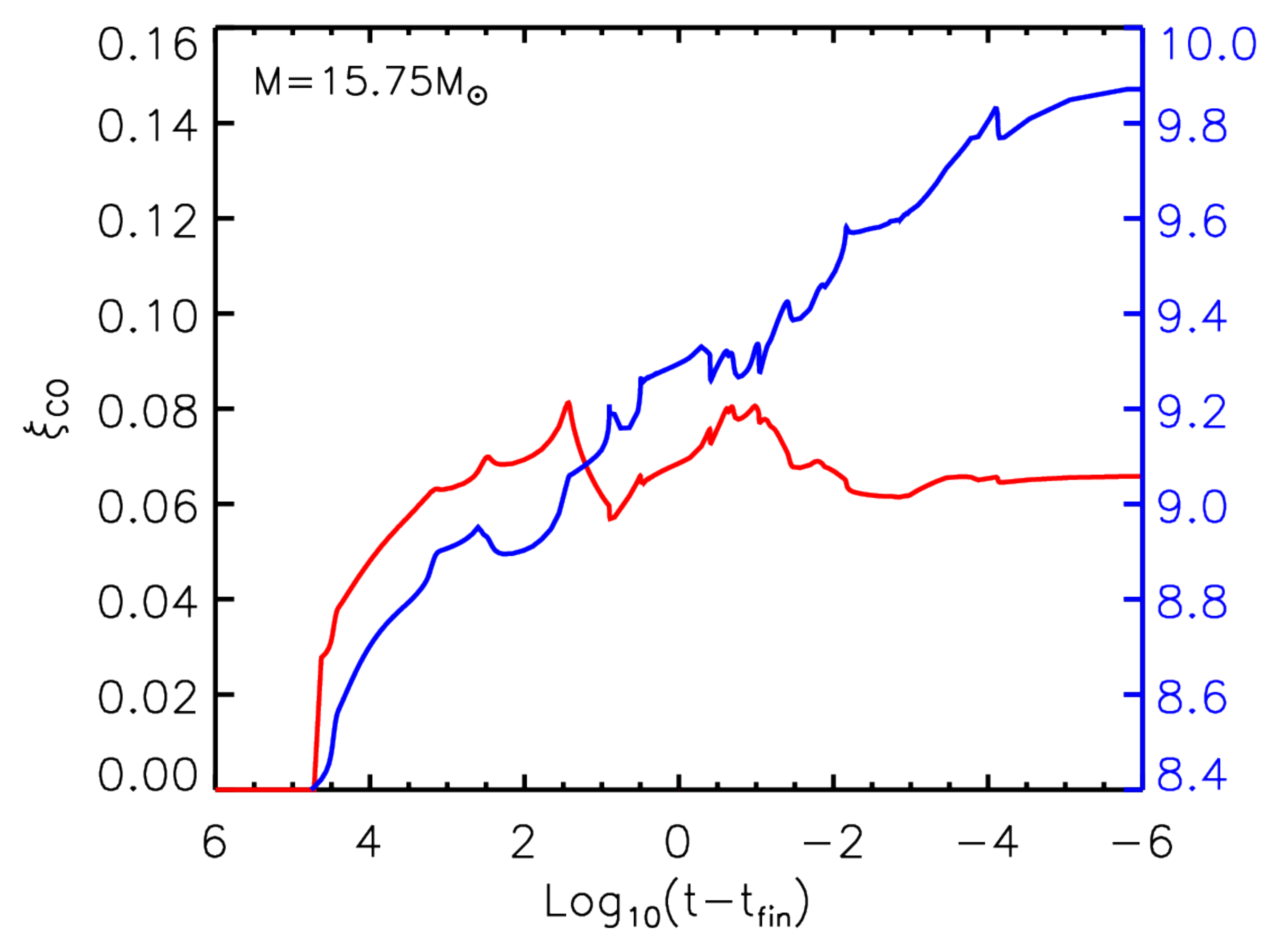}{0.33\textwidth}{(f)}}
\caption{Upper panels: Kippenhahn diagrams of the 12, 15.70 and 15.75\msun. The red, cyan and green areas mark, respectively, the convective core, the convective envelope and the convective shells. The blue line, when present, refer to the current mass of the star. Time is counted from the collapse time. Lower panels: temporal run of the central temperature (blue) and of compactness of the CO core mass (red). Each lower panel refer to the same mass  shown in its corresponding upper panel. \label{fig:kip12}}
\end{figure*}
For sake of clarity let us remind that the advancing in mass of the C burning shell is characterized by the formation of a few (usually two/three in this mass interval) convective shell episodes. The growth of these thermal instabilities has two major effects: on one side they halt (or at least slow down) the advancing of the burning shell because they continuously feed it with fresh fuel (until the convective region is rich of fuel) and, on the other side, they determine a more or less effective expansion of part of the overlying layers softening therefore their compactness, i.e. their $\rm \xi$, until they are active.

\begin{figure}[ht!]
\epsscale{1.}
\plotone{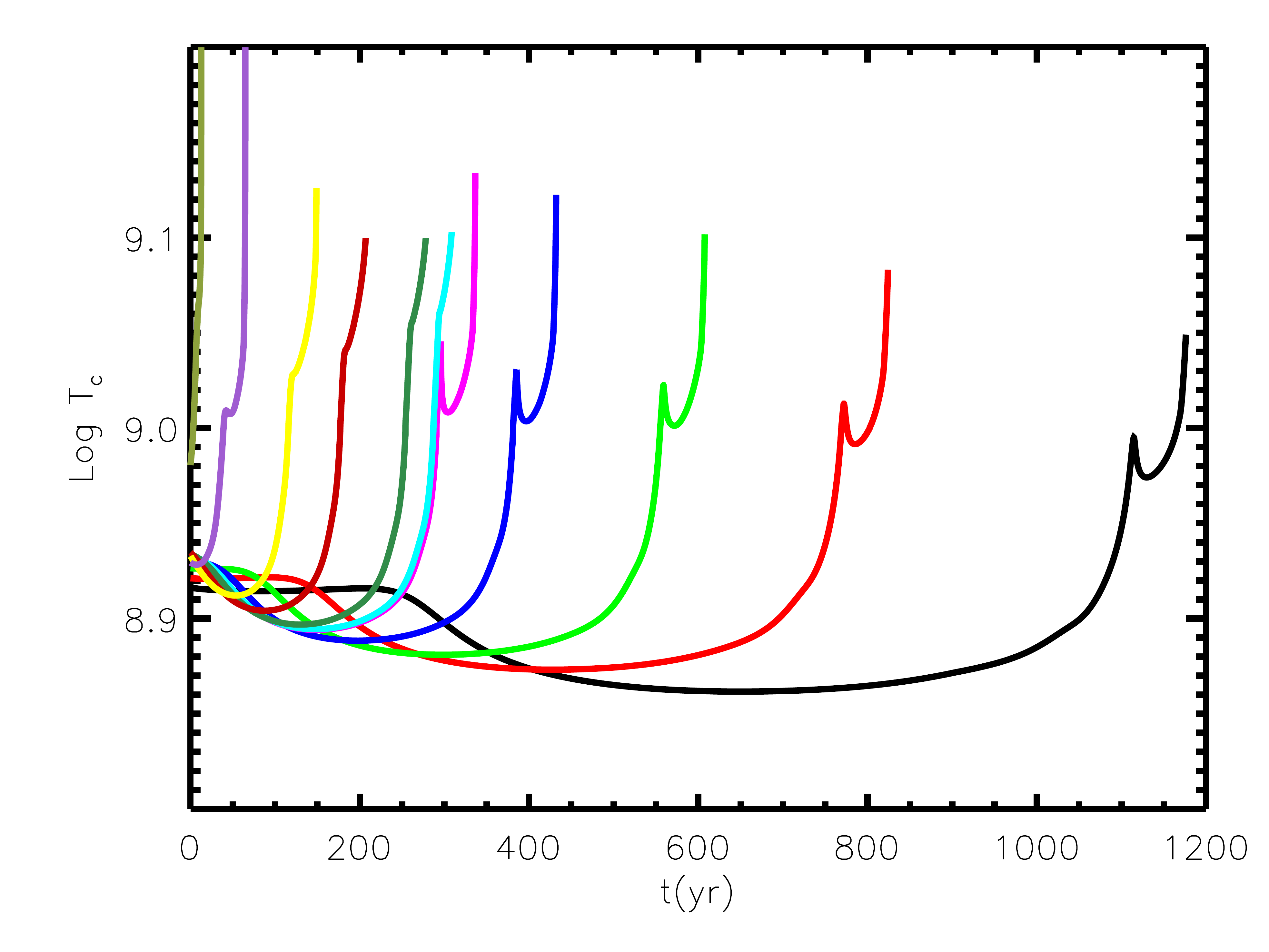}
\caption{Temporal evolution of the central temperature for a subset of models. Time starts from the end of the central He burning. The various lines refer to the 12\msun(black), 13\msun(red), 14\msun(light green), 15\msun(blue), 15.7\msun(magenta), 15.75\msun(cyan), 16\msun(dark green), 17\msun(dark red), 18\msun(yellow), 20\msun(purple) and 27\msun(grey), respectively.\label{fig:tcroc}}
\end{figure}

Figure \ref{fig:kip12} shows the Kippenhahn diagram (panel {\it a}) and the run of both $\rm \xi_{CO}$ and the central temperature (red and blue lines in panel {\it d}) of the 12\msun. A comparison between these two panels clearly shows that the formation of the convective core slows down the contraction of the core as well as its heating. The formation of the first convective shell initially leads to an expansion of the CO core ($\rm \xi_{CO}$ decreases). The same holds for the second C convective shell. Only after the exhaustion of the second convective shell the inner core is massive enough to be able to contract and heat up to the temperature necessary for the Ne photo disintegration. In fact the Ne convective core (located at $\rm Log_{10}(t-t_{fin})\sim0.8$) forms some time after the disappearance of the second C convective shell (Figure \ref{fig:kip12}). This behavior remains qualitatively unaltered up to the 15.70\msun: panels {\it b} and {\it e} in Figure \ref{fig:kip12} show the same quantities plotted for the 12\msun, but for the 15.70\msun. Above this threshold mass the evolution between the end of the central C burning and the Ne ignition changes drastically because the C exhausted core at the time of the disappearance of the first C convective shell is massive enough to contract and heat independently on the behavior of the second C burning shell. The faster contraction of the inner core forces the second C convective shell to ignite more violently than in the less massive stars and such a larger injection of energy forces the outer layers to expand, including the border of the CO core: this is the reason for the sharp decrease of $\rm \xi_{CO}$ at M=15.75\msun. Panels {\it c} and {\it f} in Figure \ref{fig:kip12} show such a change of behavior in the 15.75\msun. Figure \ref{fig:tcroc} shows, even more clearly, how the contraction timescale of the CO core changes with the initial mass. Stars in the range 12 to 15.70\msun~ show a temporary temperature decrease (a hook) during the activity of the second C convective shell while the more massive stars contract and heat without experiencing any delay in the heating of the inner core.

\begin{figure*}
\gridline{\fig{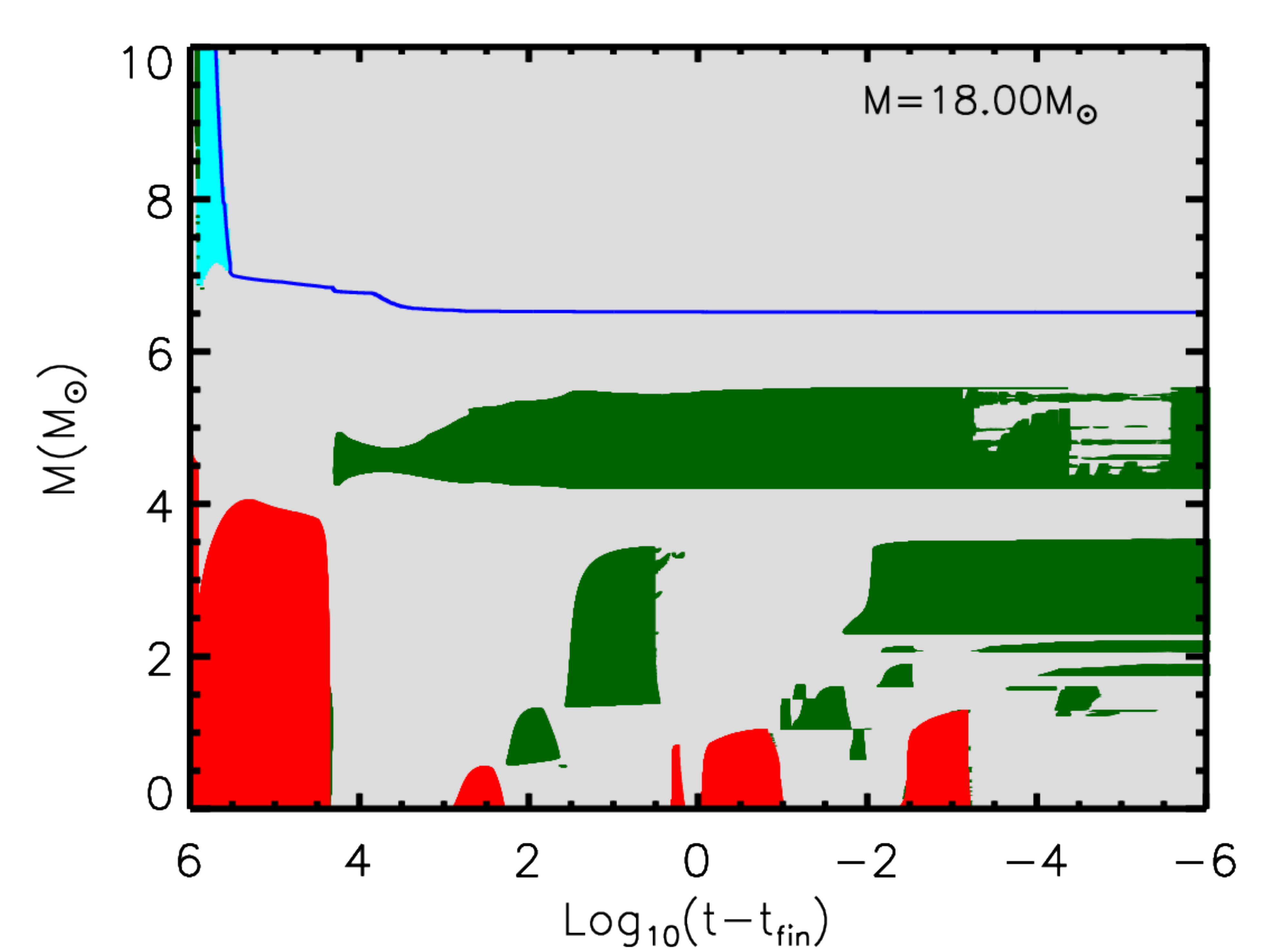}{0.3\textwidth}{(a)}
          \fig{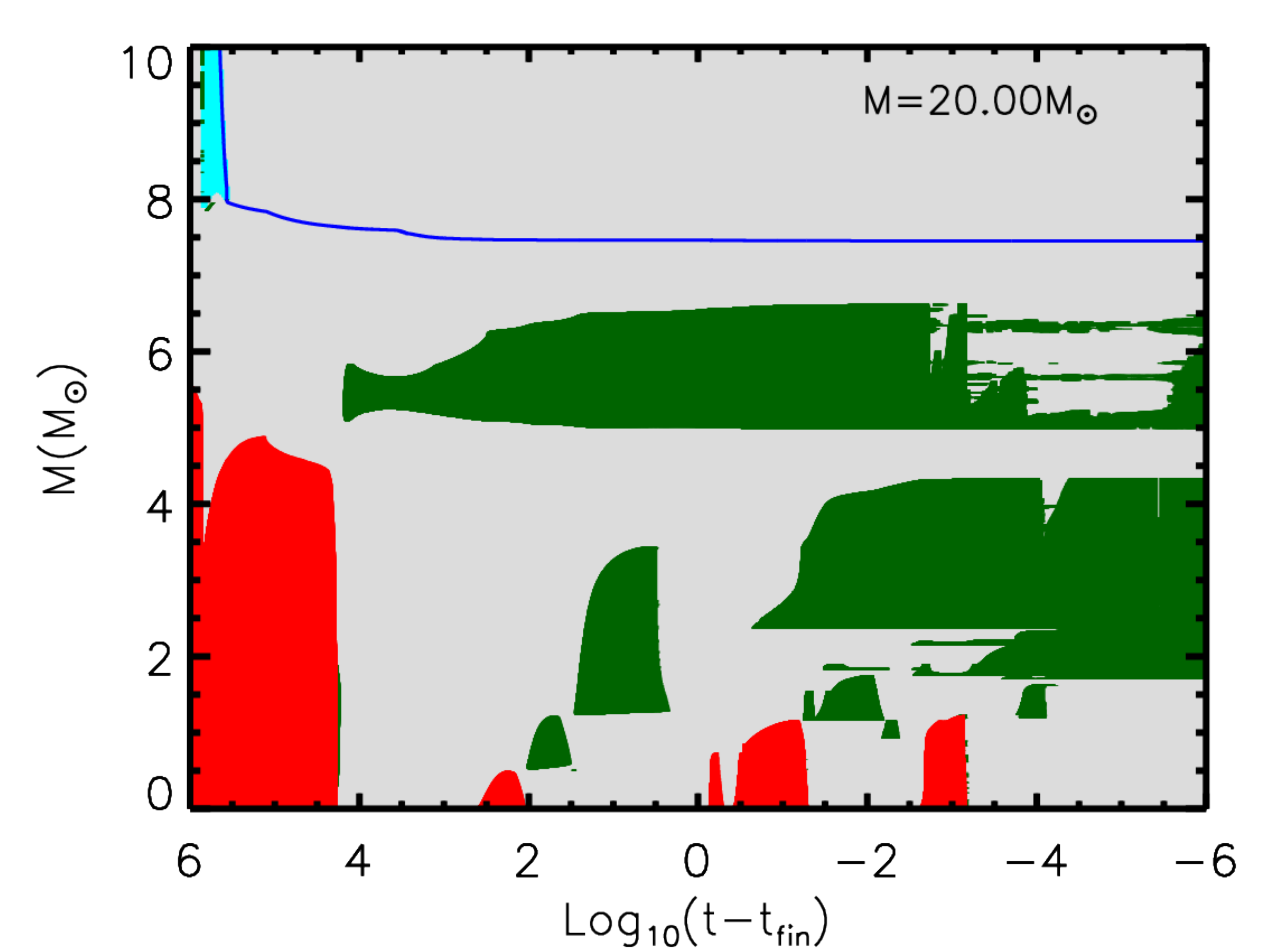}{0.3\textwidth}{(b)}
          \fig{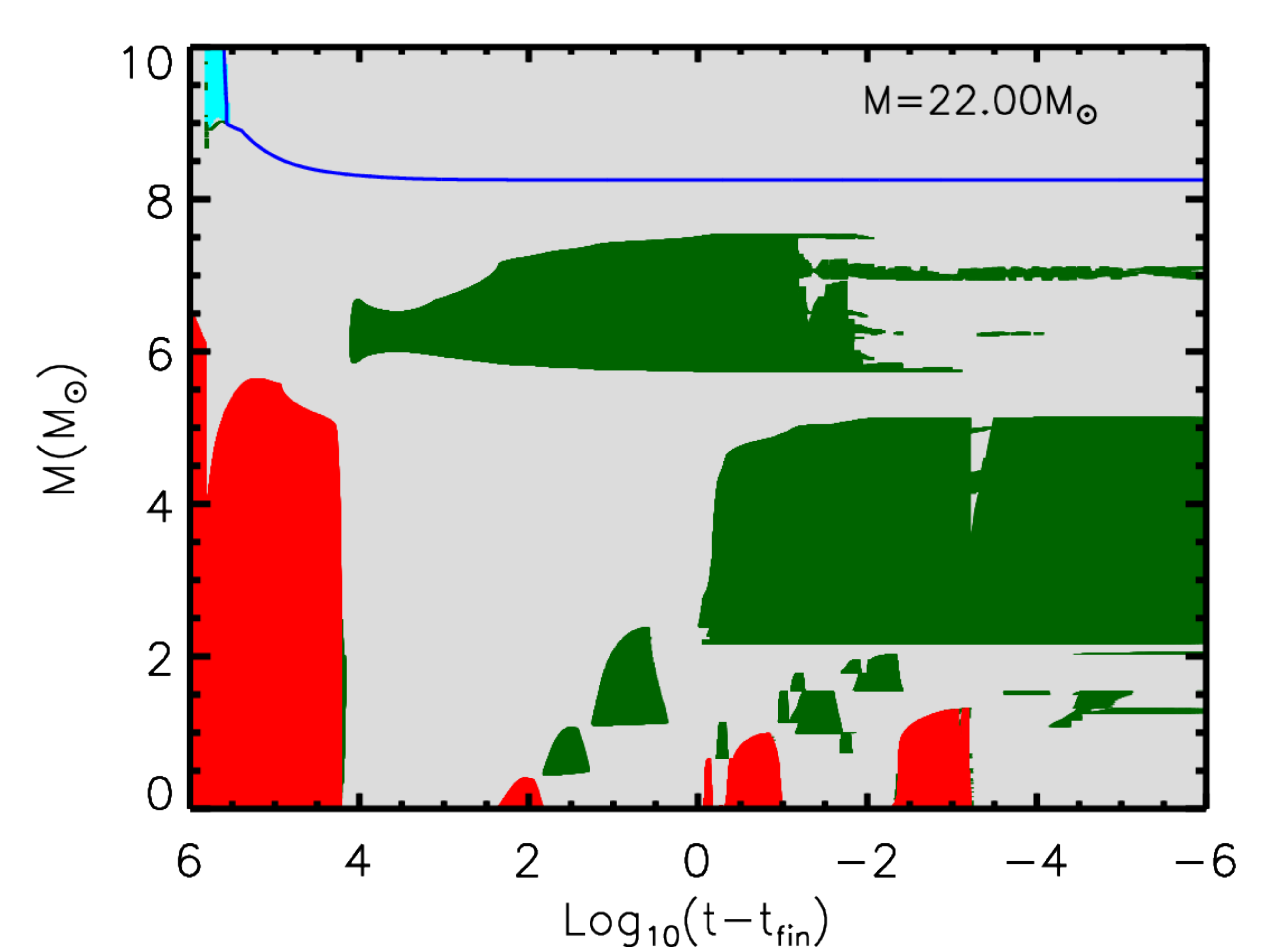}{0.3\textwidth}{(c)}}
\gridline{\fig{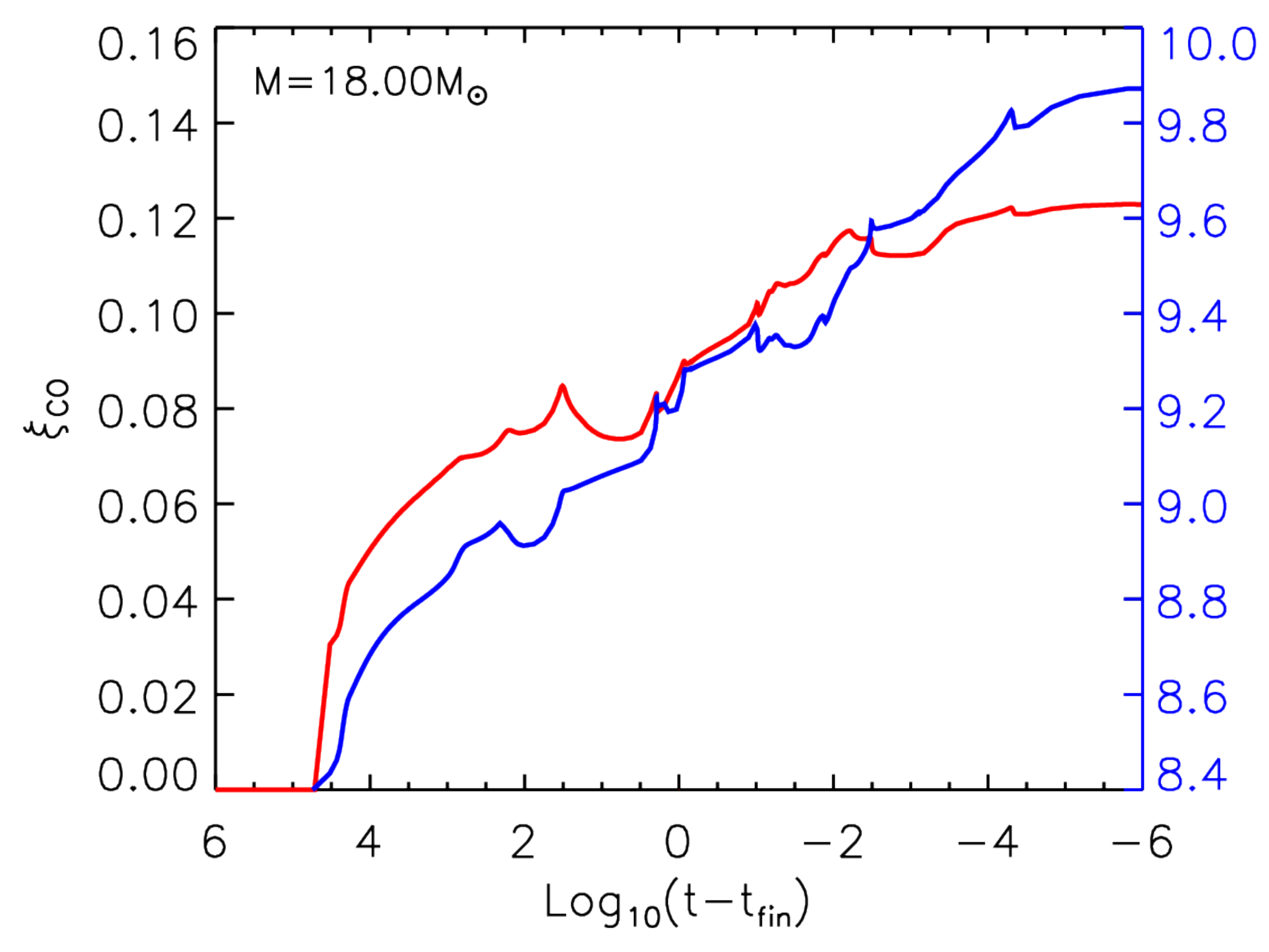}{0.33\textwidth}{(d)}
          \fig{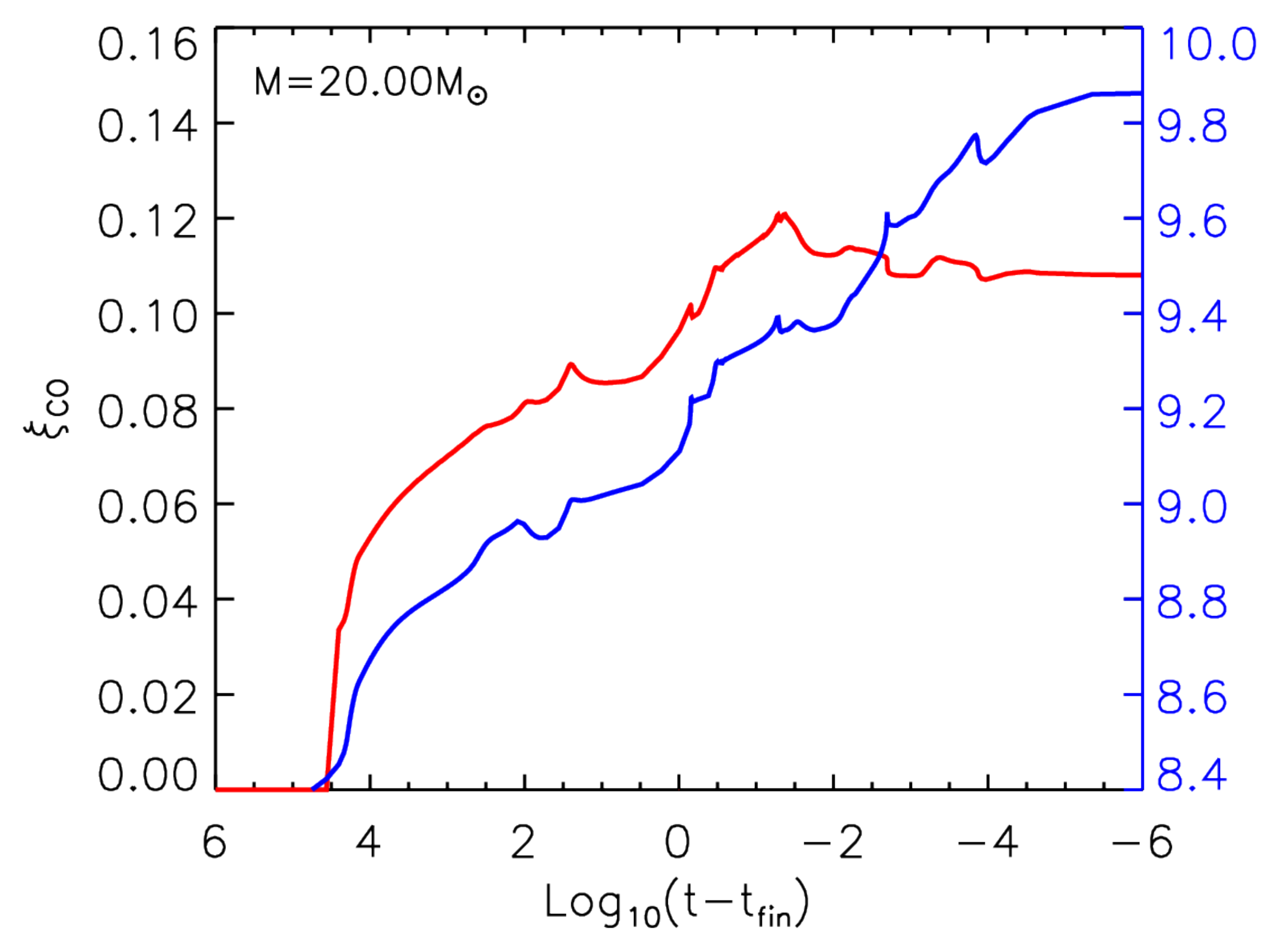}{0.33\textwidth}{(e)}
          \fig{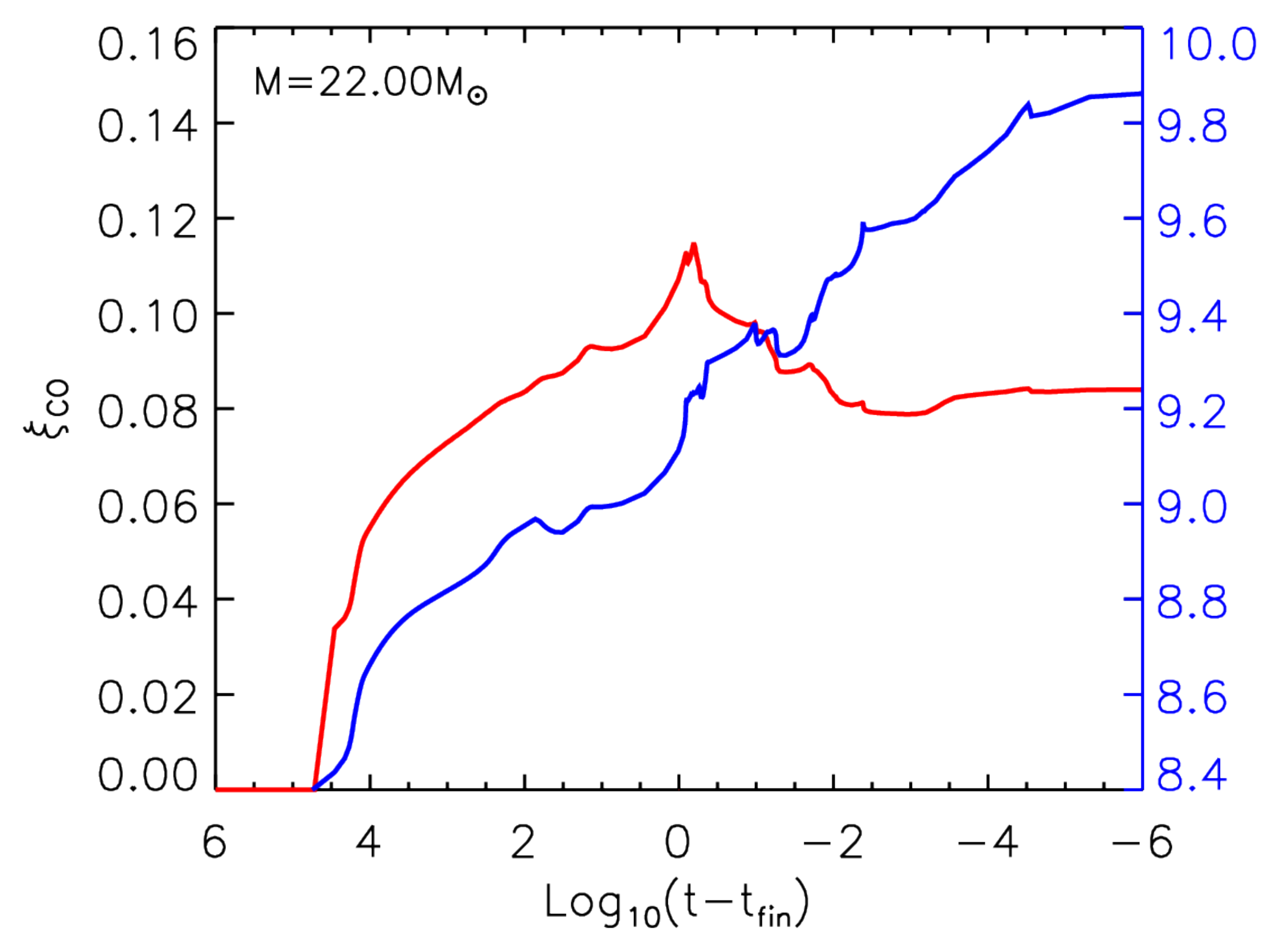}{0.33\textwidth}{(f)}}
\caption{Same as in Figure \ref{fig:kip12}.\label{fig:kip1822}}
\end{figure*}

\begin{figure*}
\gridline{\fig{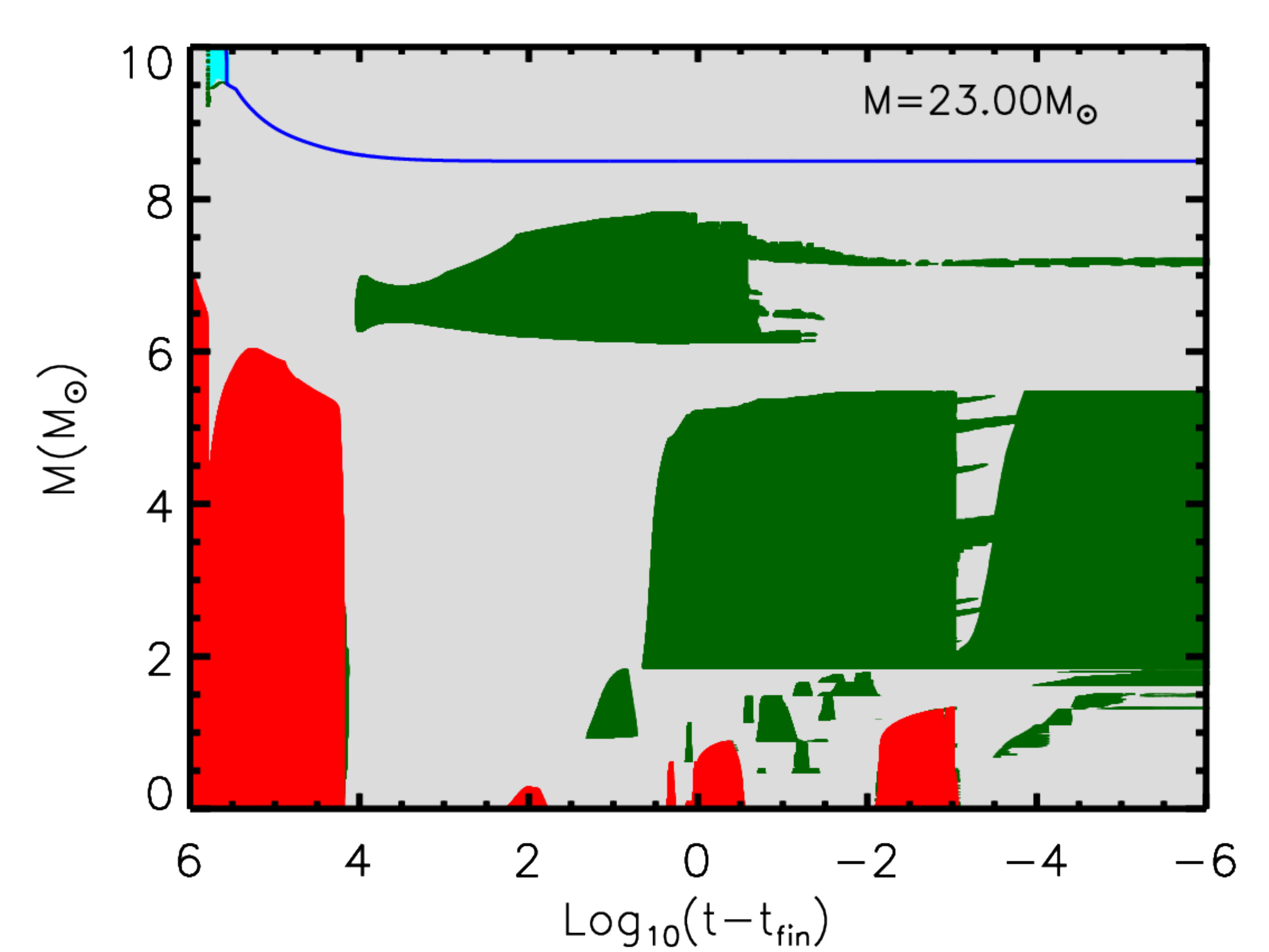}{0.3\textwidth}{(a)}
          \fig{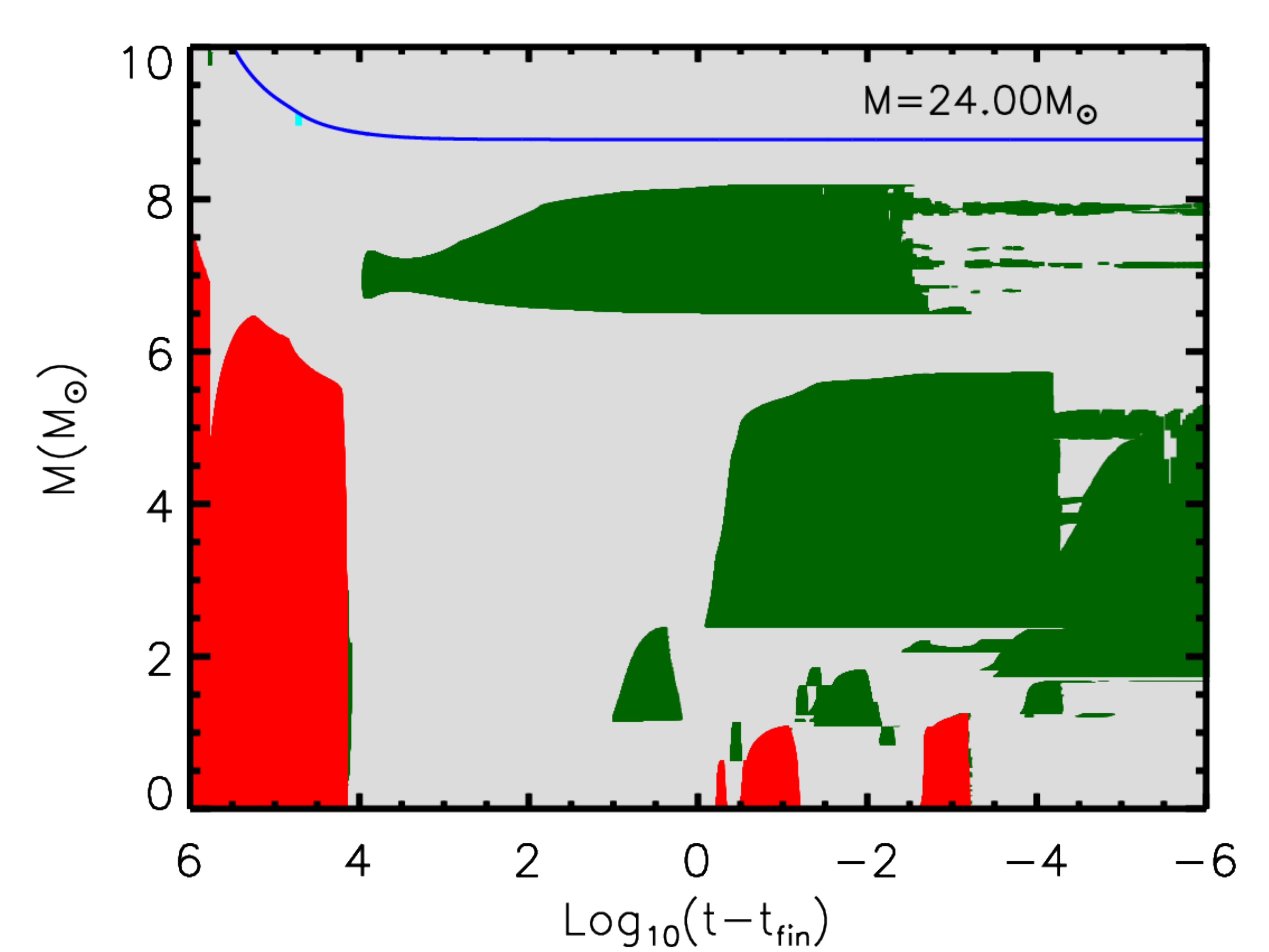}{0.3\textwidth}{(b)}}
\gridline{\fig{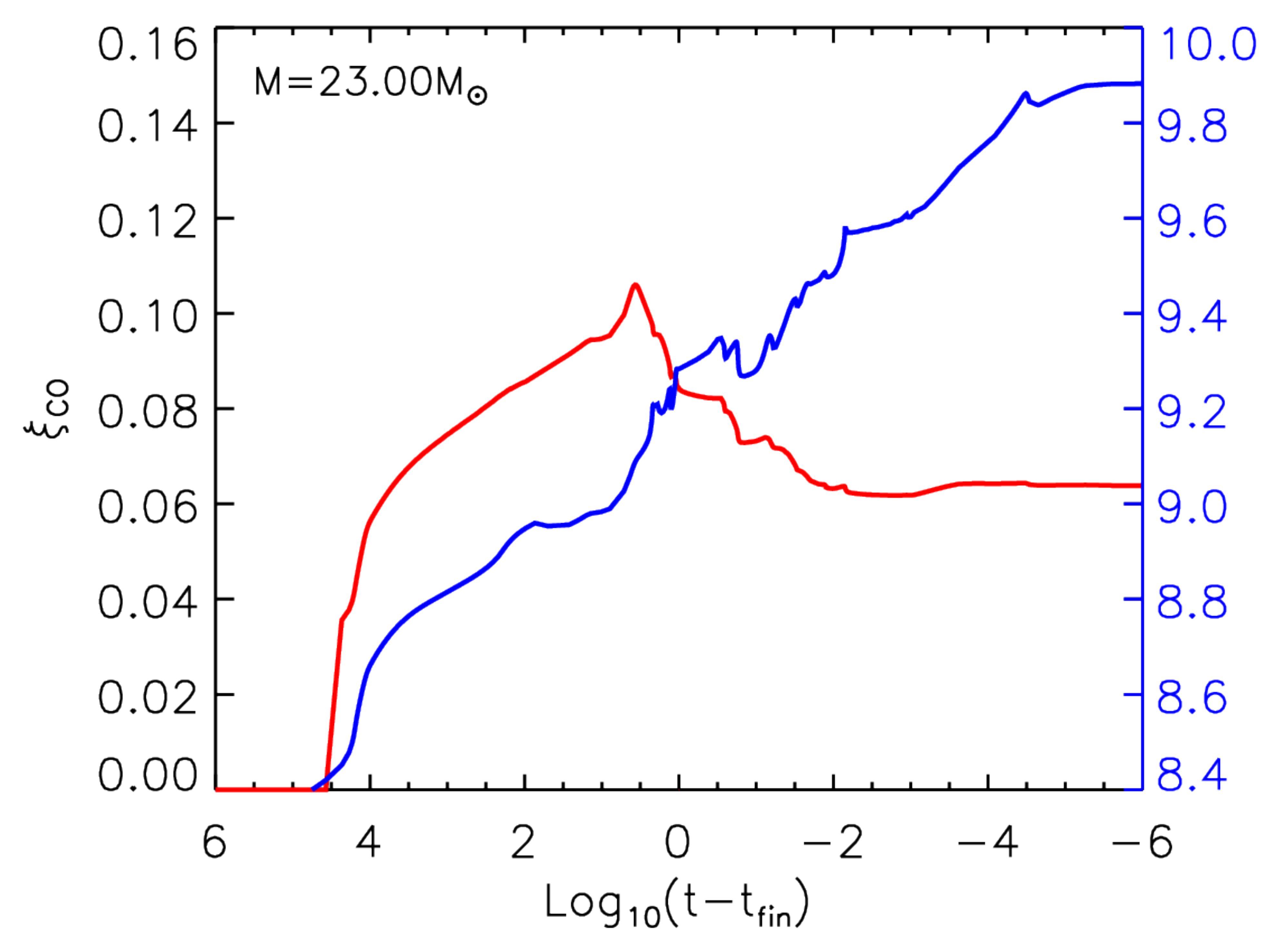}{0.33\textwidth}{(c)}
          \fig{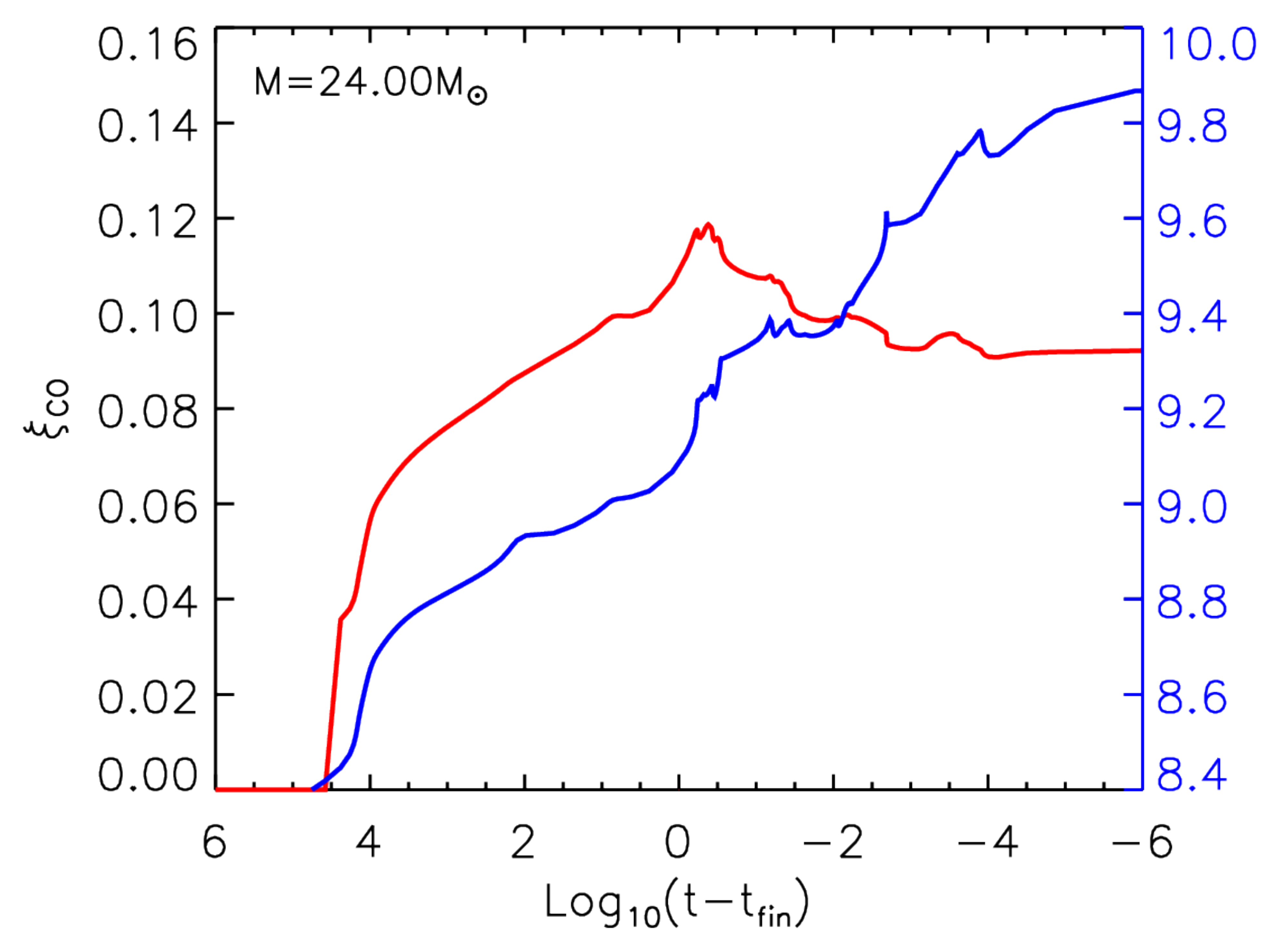}{0.33\textwidth}{(d)}}
\gridline{\fig{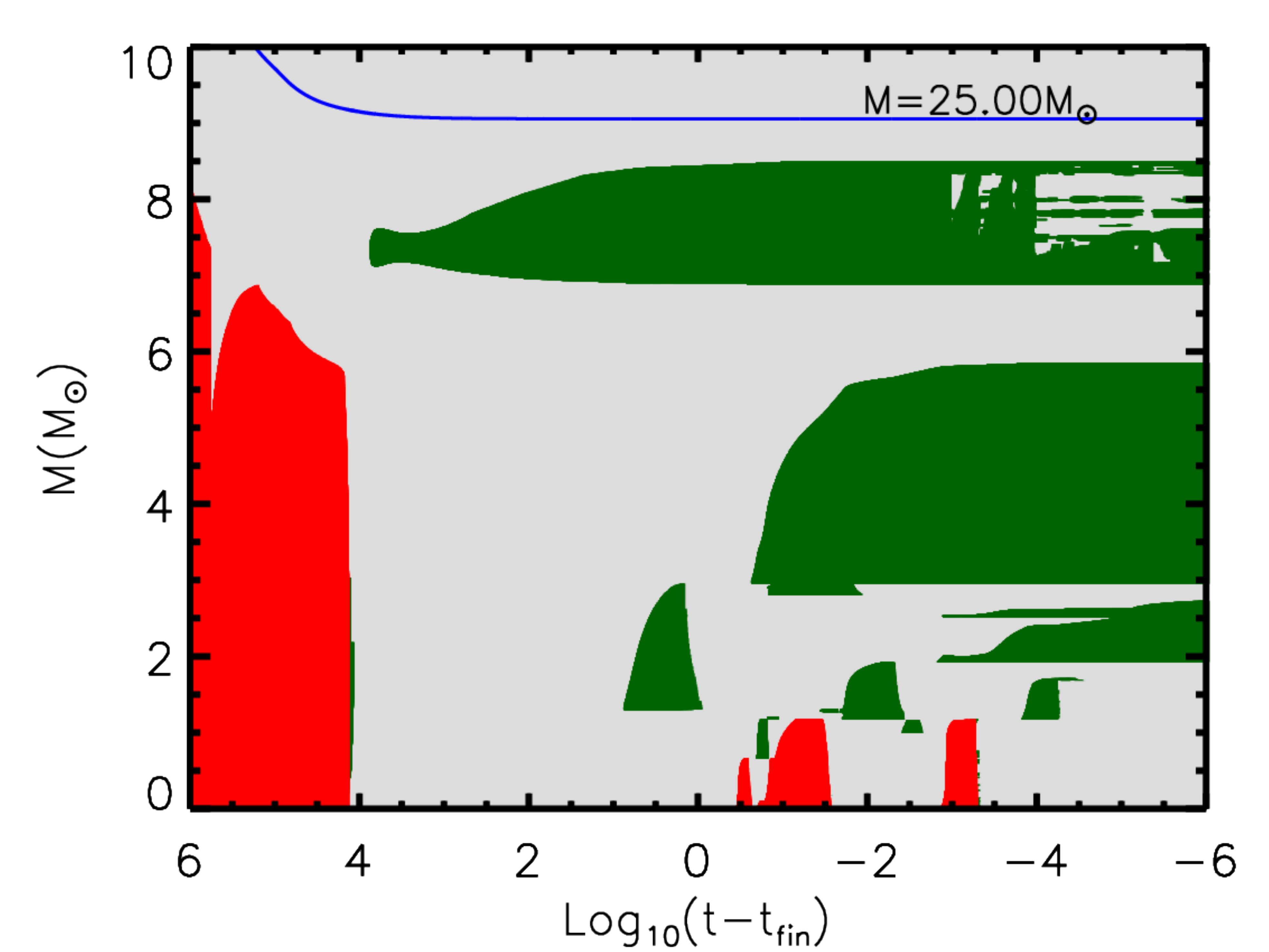}{0.3\textwidth}{(e)}
          \fig{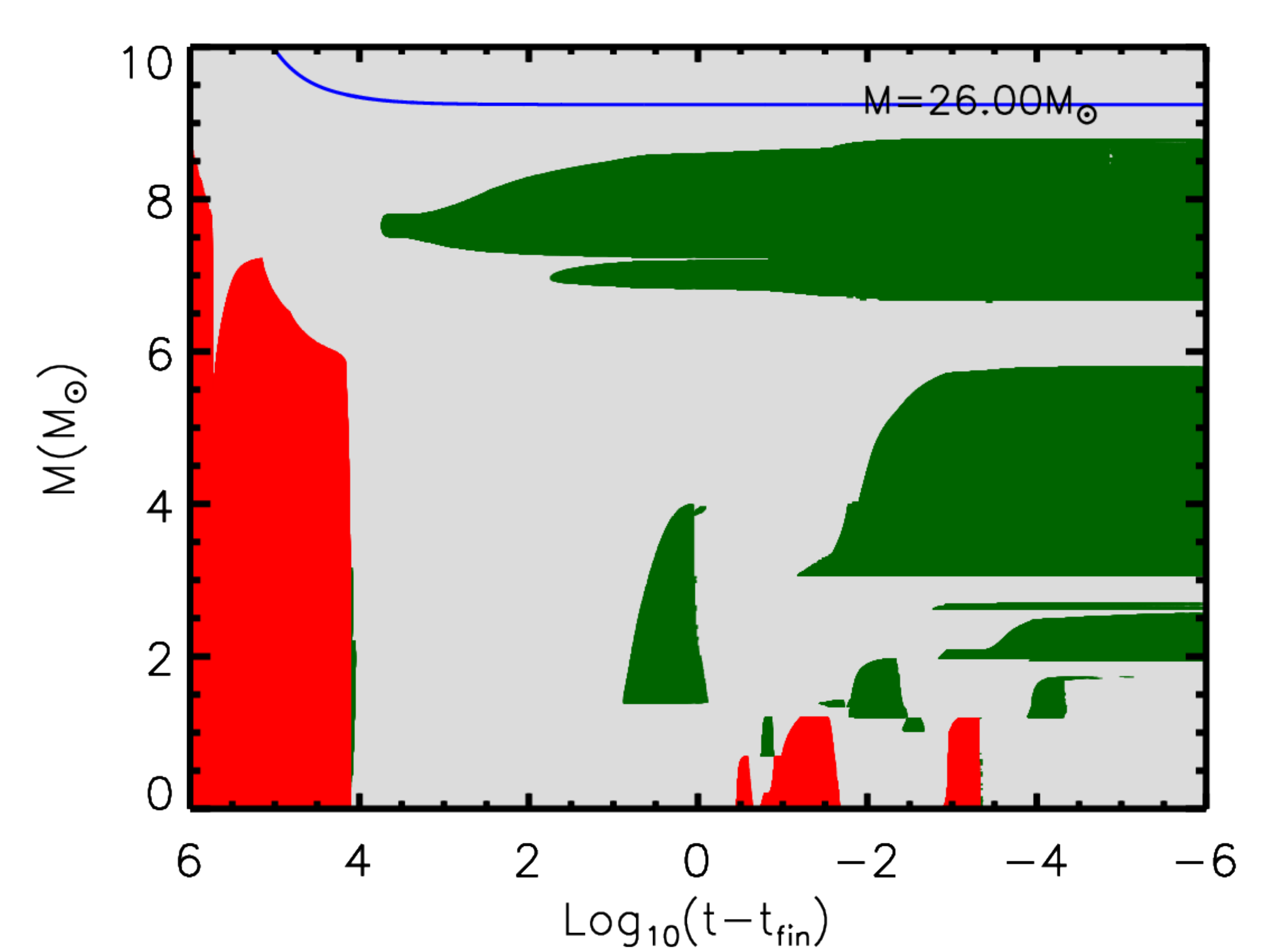}{0.3\textwidth}{(f)}}
\gridline{\fig{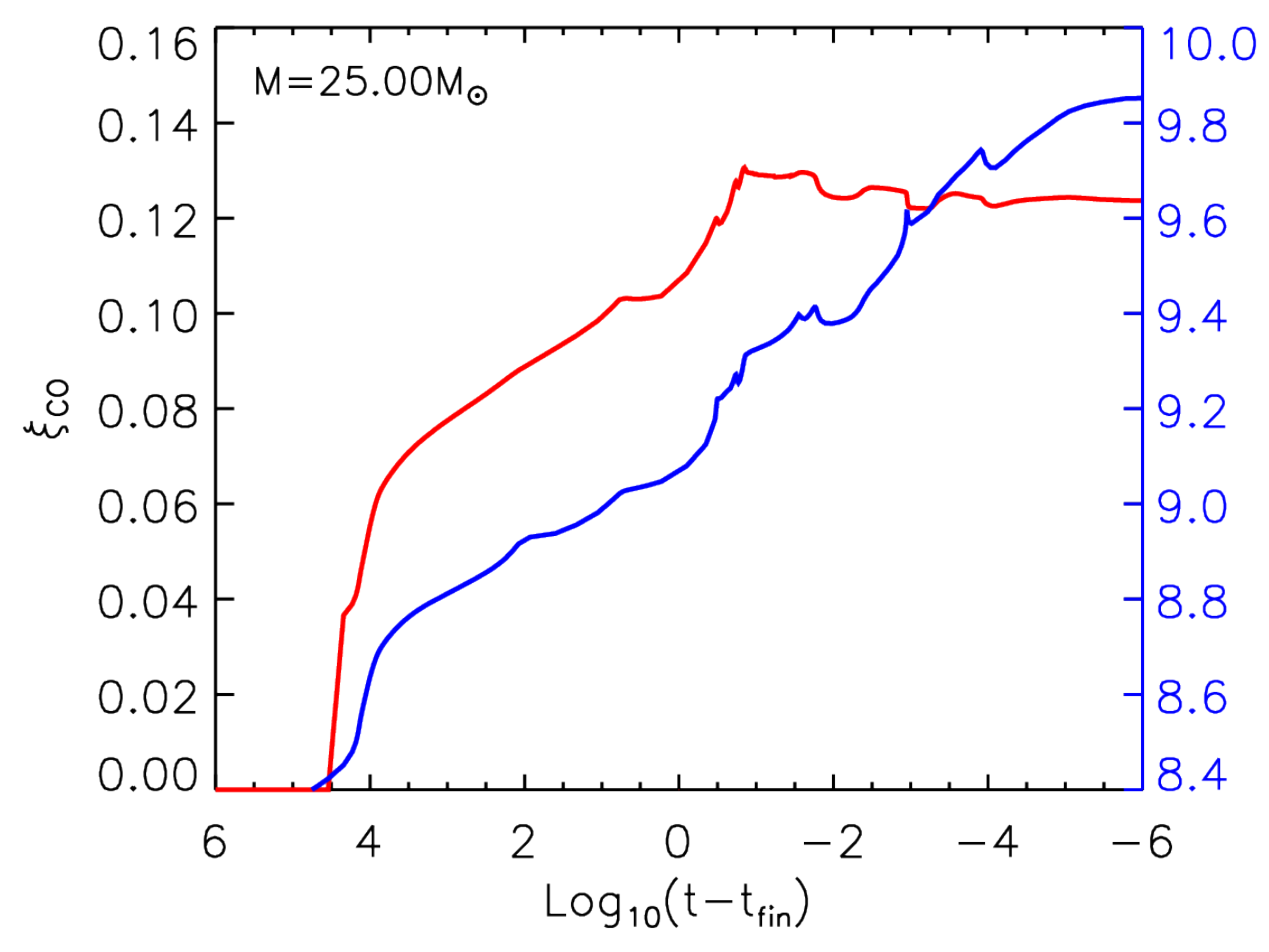}{0.33\textwidth}{(g)}
          \fig{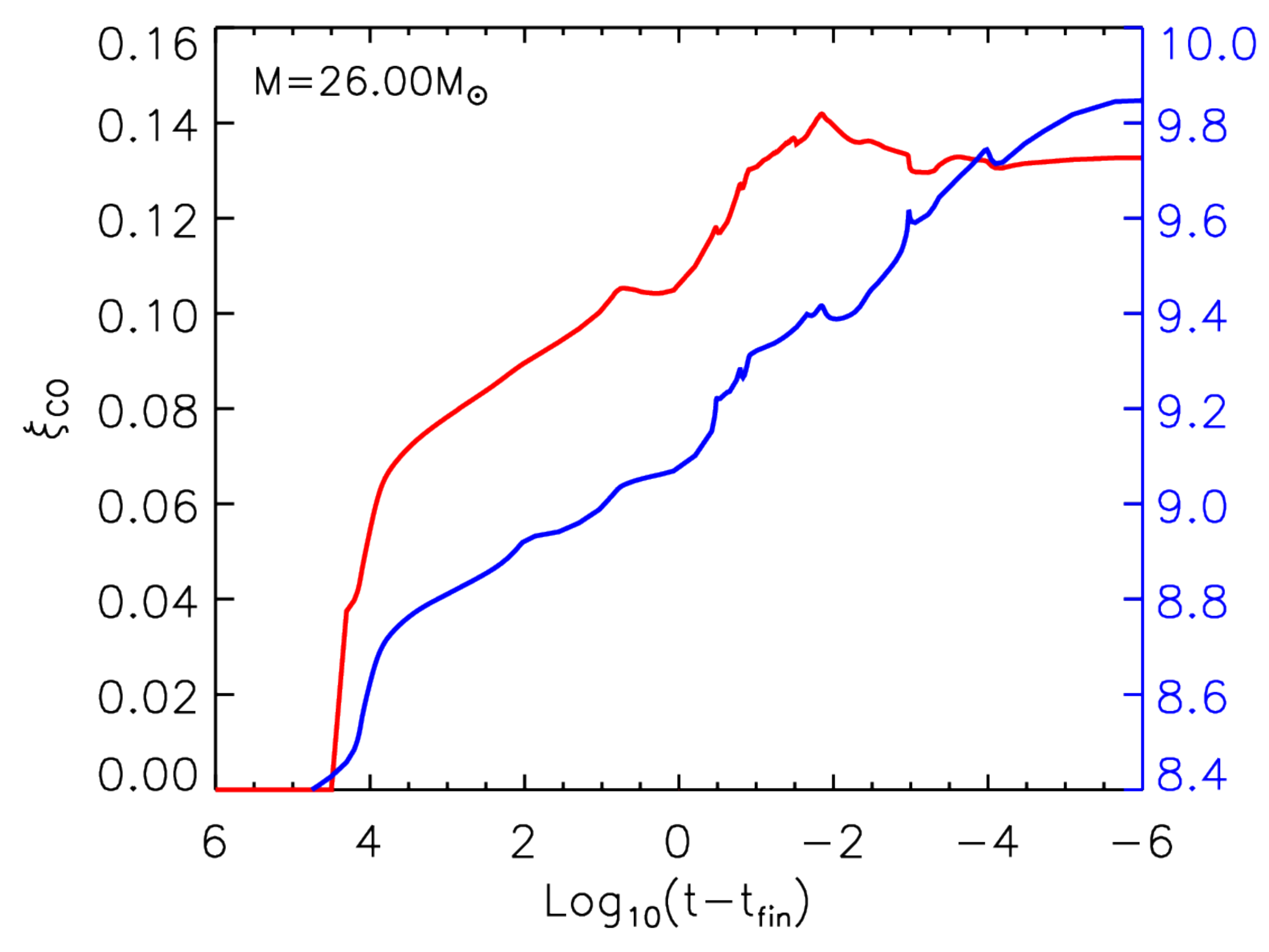}{0.33\textwidth}{(h)}}
\caption{Same as in Figure \ref{fig:kip12}.\label{fig:kip2327}}
\end{figure*}

Stars in the range 15.75\msun~ and roughly 17\msun~ reach the Ne ignition with a $\rm \xi_{CO}$ smaller (i.e. a CO core more expanded) than the one they had at the end of the central C burning because of the power of the second convective shell. But, as the initial mass increases, the second C convective shell weakens progressively and it even vanishes before Ne ignites, so that the CO core has time to further contract by the time the center reaches the condition for the Ne burning. The net consequence is a progressive increase of $\rm \xi_{CO}$. This effect is readily visible in Figure \ref{fig:kip1822}, where the Kippenhahn diagrams of the 18, 20 and 22\msun~ are shown together to the temporal evolution of both $\rm \xi_{CO}$ and central temperature: the size of the second C convective shell progressively reduces moving from the 18 to the 22\msun~ while the Ne ignition shifts towards later times with respect to the end of the second C convective shell. Above $\sim22$\msun~ $\rm \xi_{CO}$ inverts its trend with the initial mass: the responsible for this turn down is the early formation of the third C convective shell. Up to now we have not mentioned the third convective shell because it forms after the Ne burning in masses smaller than $\sim22$\msun. The systematic decrease of the power of the second C convective shell as the initial mass increases, speeds up the contraction and heating of the CO core so that the formation of the third convective shell progressively anticipates in time and around the 22\msun~ its formation almost coincides with the Ne ignition. Similarly to what happens around the 15.7\msun, the growth of this convective shell forces the overlying layers to expand and hence $\rm \xi_{CO}$ to decrease. The right panels in Figure \ref{fig:kip1822} show that at the beginning of the Ne burning ($\rm Log(t-t_{\rm end})\sim-0.02$) $\rm \xi_{CO}$ begins to drop because of the growth of the third convective shell. In the mass range 22 to 22.9\msun~ the third C convective shell systematically forms before the central Ne ignition and this occurrence leads to a progressive decrease of $\rm \xi_{CO}$ in this mass interval. As the initial mass continues to increase (above $\sim22.9$\msun) also the strength of the third C convective shell progressively weakens and, accordingly, $\rm \xi_{CO}$ increases again. The behavior of the third C convective shell is well summarized in Figure \ref{fig:kip2327} where the same quantities plotted for the less massive stars are now shown for the 23, 24, 25 and 26\msun.

The cyan dots in Figure \ref{fig:csico} show the trend of $\rm \xi_{CO}$ at the central Si exhaustion. It is worth noting that the main features already present at the Ne ignition are still there, i.e. the discontinuity at 15.75\msun~ and the minimum at 22.8\msun. In addition to this, it is worth noting that while the CO core of stars in the intervals 12-20\msun~ and 25-27.95\msun~ shows a more compact structure with respect to the one they have at the central Ne ignition (because they tend on average to contract as the center evolves), stars in the range 20 to 25\msun~ show an opposite behavior, reaching the end of the central Si burning with a CO core more expanded than at the central Ne ignition: the reason is that this is the mass interval in which the third C convective shell reaches its maximum strength and extension and we have already seen before that a very strong burning shell forces the overlying layers to expand and hence to reduce their compactness. The lower panels in Figures \ref{fig:kip12}, \ref{fig:kip1822} and \ref{fig:kip2327} clearly show that the compactness of the CO core does not increase any more (but it can decrease) after the formation of the last C convective shell. The small drop (and scatter) in $\rm \xi_{CO}$ that is present at $\sim26$\msun~ in Figure \ref{fig:csico} is due to the formation of a small He convective shell (in the tail of the He profile) that merges with the main one. The sudden shift of the base of the new wider He convective shell to a more internal mass coordinate obviously forces also a jump of $\rm \xi_{CO}$. The blue dots in Figure \ref{fig:csico} show the final compactness of the CO cores of our models at the onset of the core collapse. With respect to the end of the central Si burning there is now only a modest or even negligible variation of the compactness of the CO core mass. A last thing worth noting is that even if the trend of $\rm \xi_{CO}$ with the initial mass is not monotonic, the correlation is extremely tight, there is basically no scatter of the points (no chaotic behavior) around the average trend line.

\begin{figure}[ht!]
\epsscale{1.}
\plotone{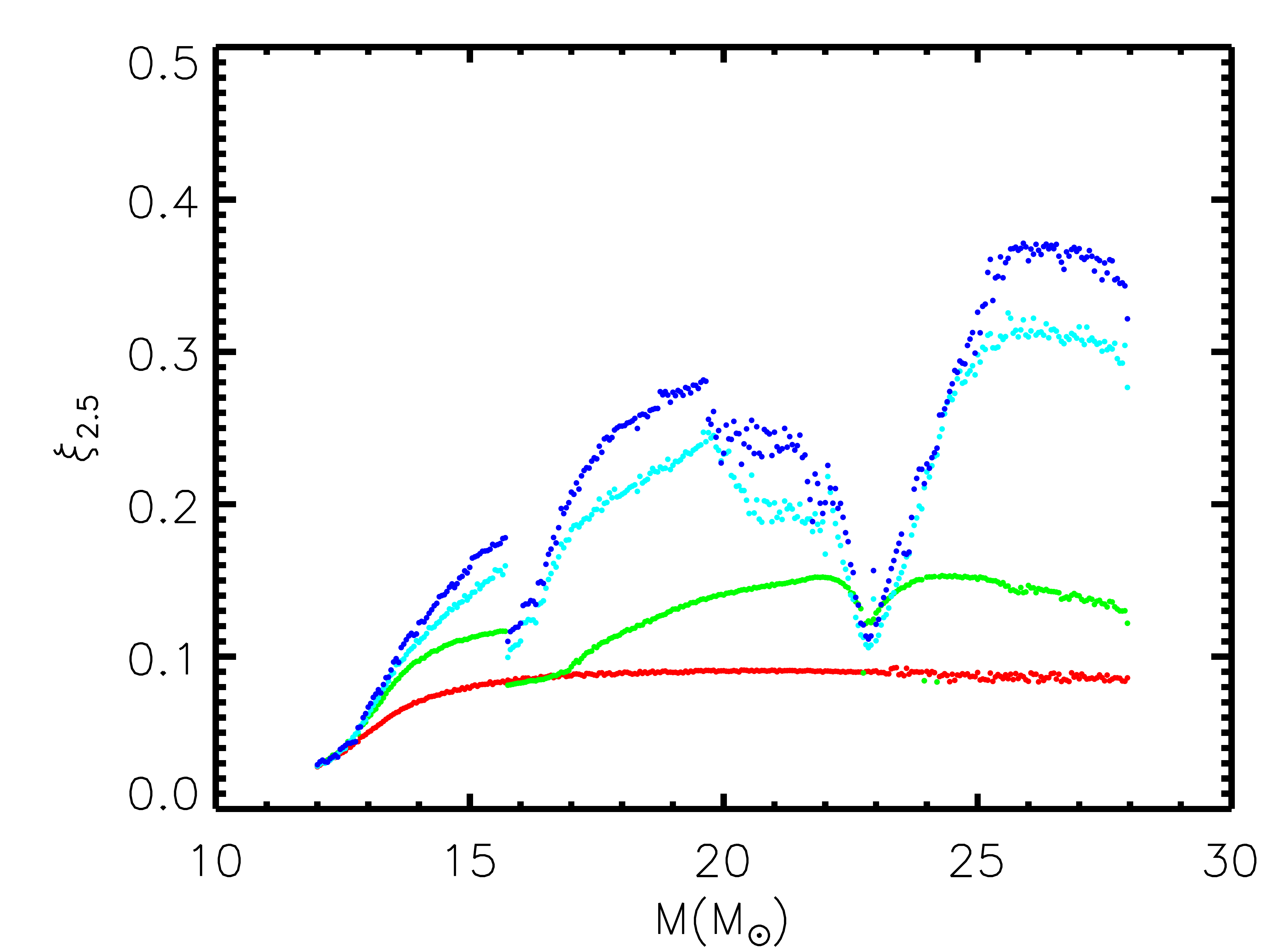}
\caption{Compactness of the mass coordinate 2.5\msun~ at various phases: end of the central C burning (red), central Ne ignition (green), central Si exhaustion (cyan) and last model (blue). \label{fig:csimini}}
\end{figure}

The second mass location that is worth discussing is the one corresponding to 2.5\msun. The reason is that this mass location has been used \citep{oo11,oo13,sw14,su18}  as a proxy for the explodability of a model. Though we do not discuss in this paper the connection between compactness and explodability, we think to be interesting to show and discuss the compactness of this layer that, in a large fraction of the models in the present grid ($\rm 14.00\leq M(M_\odot)\leq 24.25$), is located within the last, most extended, C convective shell. Figure \ref{fig:csimini} shows the run of $\rm \xi_{2.5}$ at some selected phases: the end of the central C burning (red dots), the beginning of the Ne photo disintegration (green dots), the end of the central Si burning (cyan dots) and the last model (blue dots). All the trends plotted in this figure show features that are strongly related to the ones already discussed for $\rm \xi_{CO}$ (Figure \ref{fig:csico}) and therefore also them are tied to the behavior of the C burning shell. The scaling with the initial mass is still clean up to the end of the central C burning, while the various features begin to appear in the passage from the end of the central C burning to the Ne ignition. The evolution beyond the Ne burning amplifies the features already present at central Ne ignition.  The discontinuity present at $\sim$20\msun~ at the onset of the collapse marks the minimum mass in which a powerful third C convective shell forms (central panels in Figure \ref{fig:kip1822}).

\begin{figure}[ht!]
\epsscale{1.}
\plotone{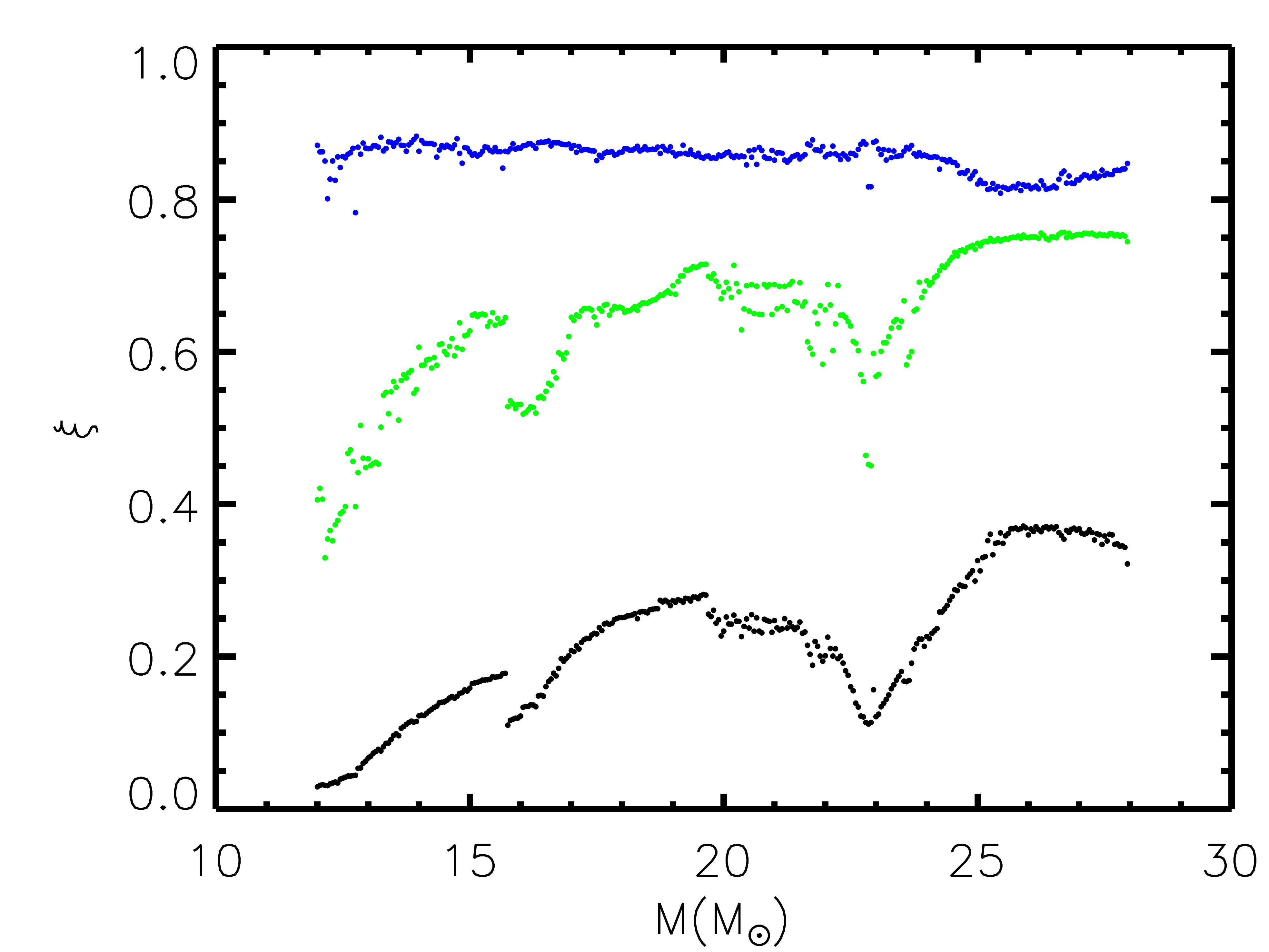}
\caption{Final compactness of the models for three different mass coordinate: 2.5\msun~(black), base of the O burning shell (green) and the Fe core mass (blue).\label{fig:csiall}}
\end{figure}

The third mass location worth being presented is the compactness of the knee present in the final M-R relation. Such a knee is sculpted by the O burning shell that is located roughly at 1.7\msun ($\pm$ 0.2\msun) in the mass interval discussed in this paper and therefore we chose this mass location to determine the compactness of the knee. Figure \ref{fig:csiall} shows the run of $\rm \xi_{knee}$ (green dots) together to the $\rm \xi_{2.5}$ (black dots). Once again the main features shown by $\rm \xi_{knee}$ are the same already discussed above and this reinforces the idea that the general trend of the compactness of a star with the initial mass is dictated by the ability of the C burning in forming powerful convective shells and in advancing in mass.

There is however a third set of points in Figure \ref{fig:csiall}. The blue dots show the trend of $\rm \xi_{1.5}$, i.e. the compactness of a layers that represents fairly well the average location of the Fe core of the present set of models. In this case there is practically no trend with the initial mass and this is due to the fact that towards the end of their hydrostatic evolution massive stars tend to share a similar M-R relation behind the Si shell.

\section{Comparison with similar computations} \label{sec:comp}
The scaling of the compactness of the massive stars with the initial mass has been discussed in the literature in several papers (see Section \ref{sec:intro}); one of the most extensive studies on this subject published up to now is the one by \cite{su18} (hereinafter SWH18). One of the key results of that paper (already found in the previous ones of the same series) is that the final compactness of the stars shows a significant scatter around the main trend at least in some mass intervals. The authors interpret this result as an intrinsic property of these stellar models because their evolution is "statistical in nature". Given the relevance of the final compactness of a star at the onset of the core collapse because of its intimate connection to the possible success/failure of the explosion, it is useful to compare their results to ours and to briefly comment them. 

\begin{figure*}
\gridline{\fig{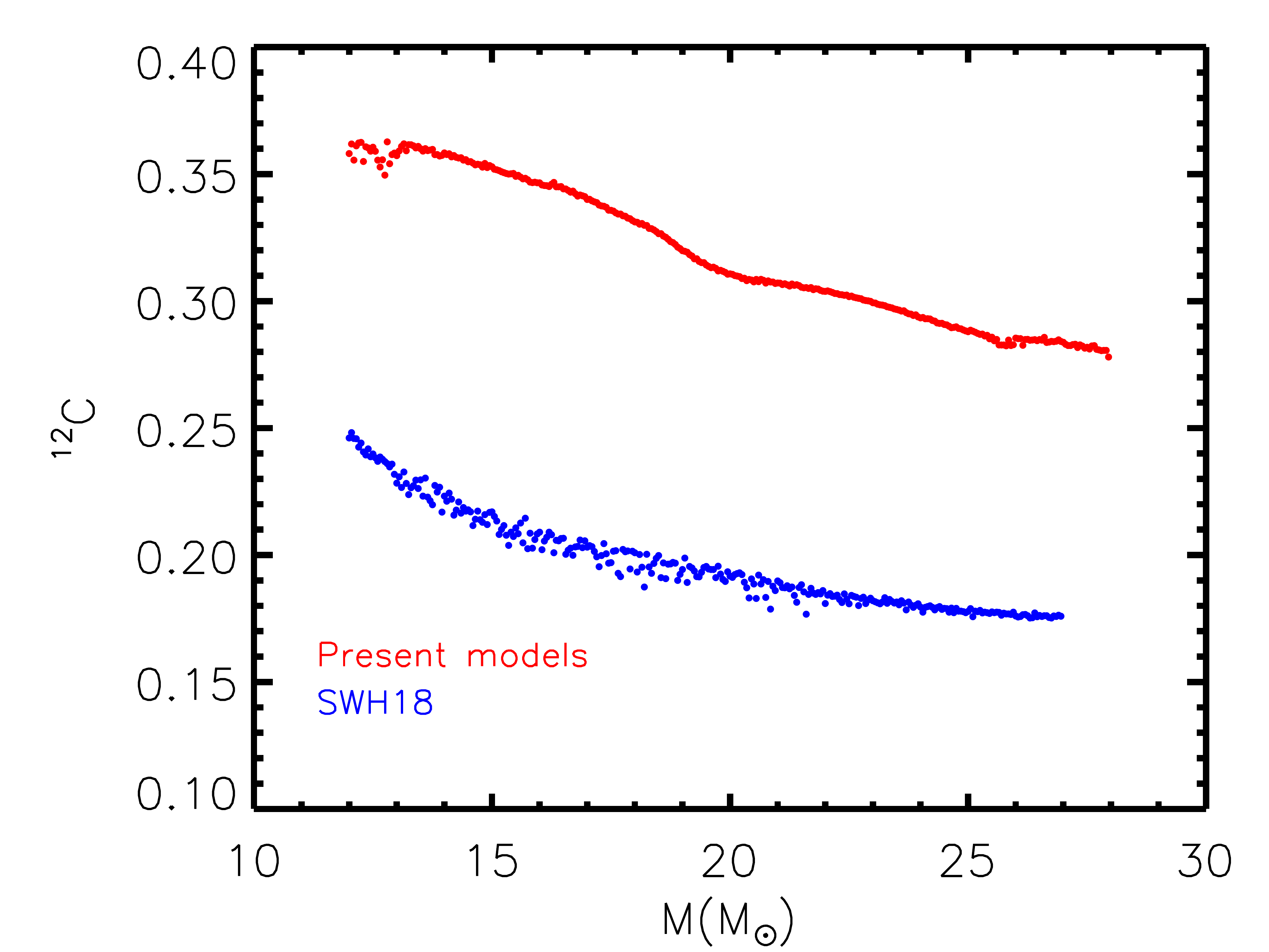}   {0.45\textwidth}{(a)}
          \fig{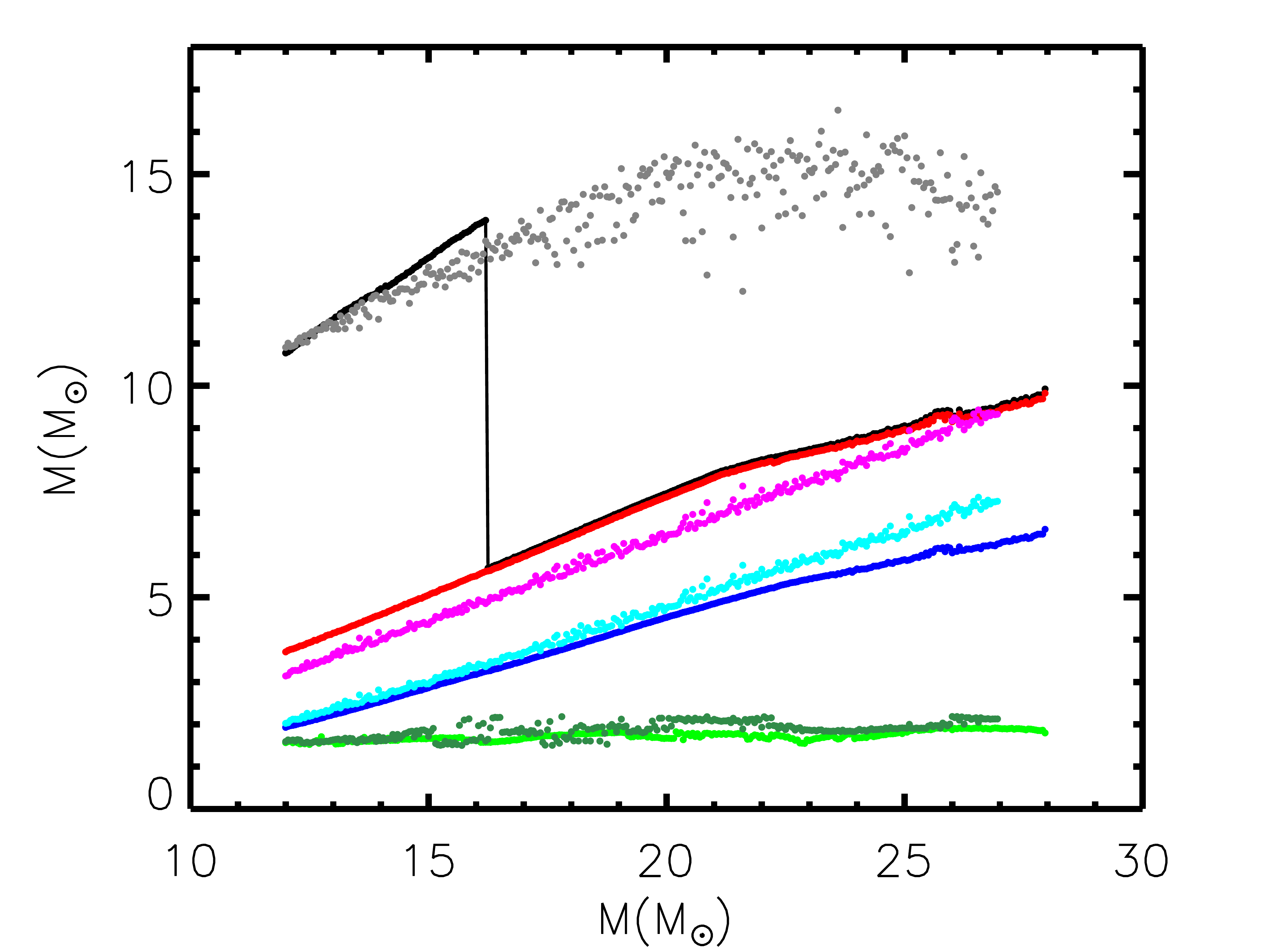}   {0.45\textwidth}{(b)}}
\gridline{\fig{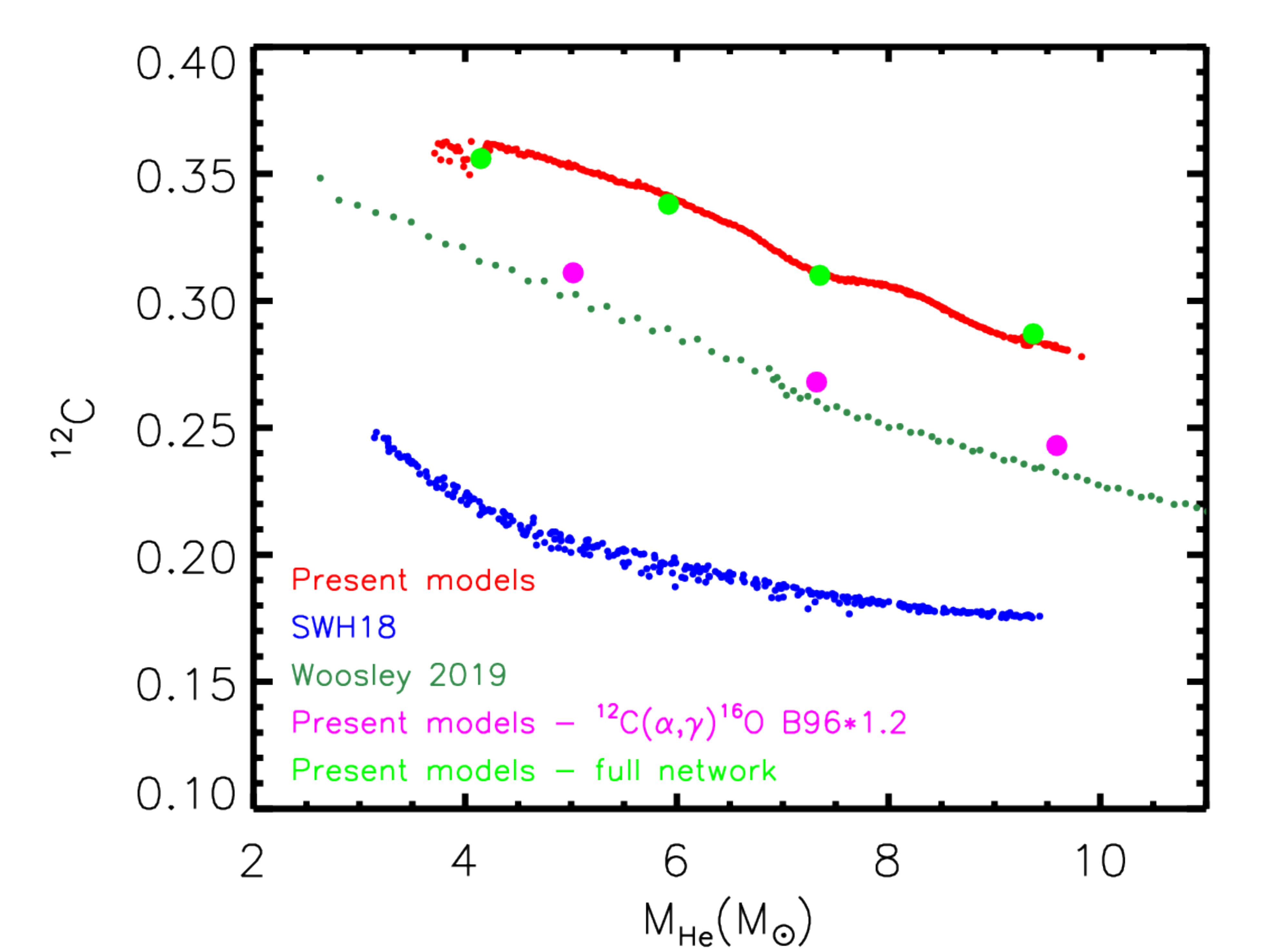}   {0.45\textwidth}{(c)}
          \fig{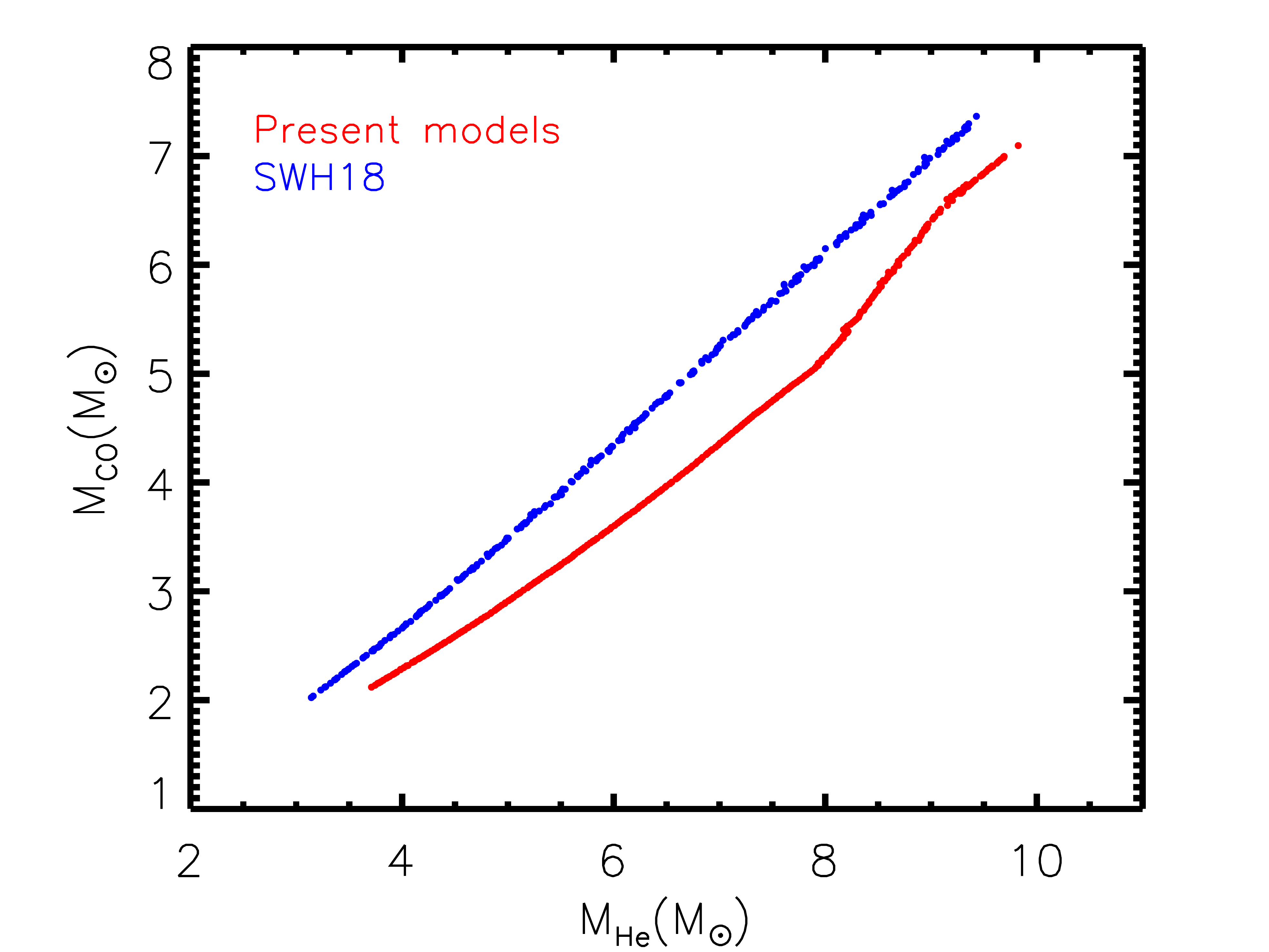}   {0.45\textwidth}{(d)}}
\gridline{\fig{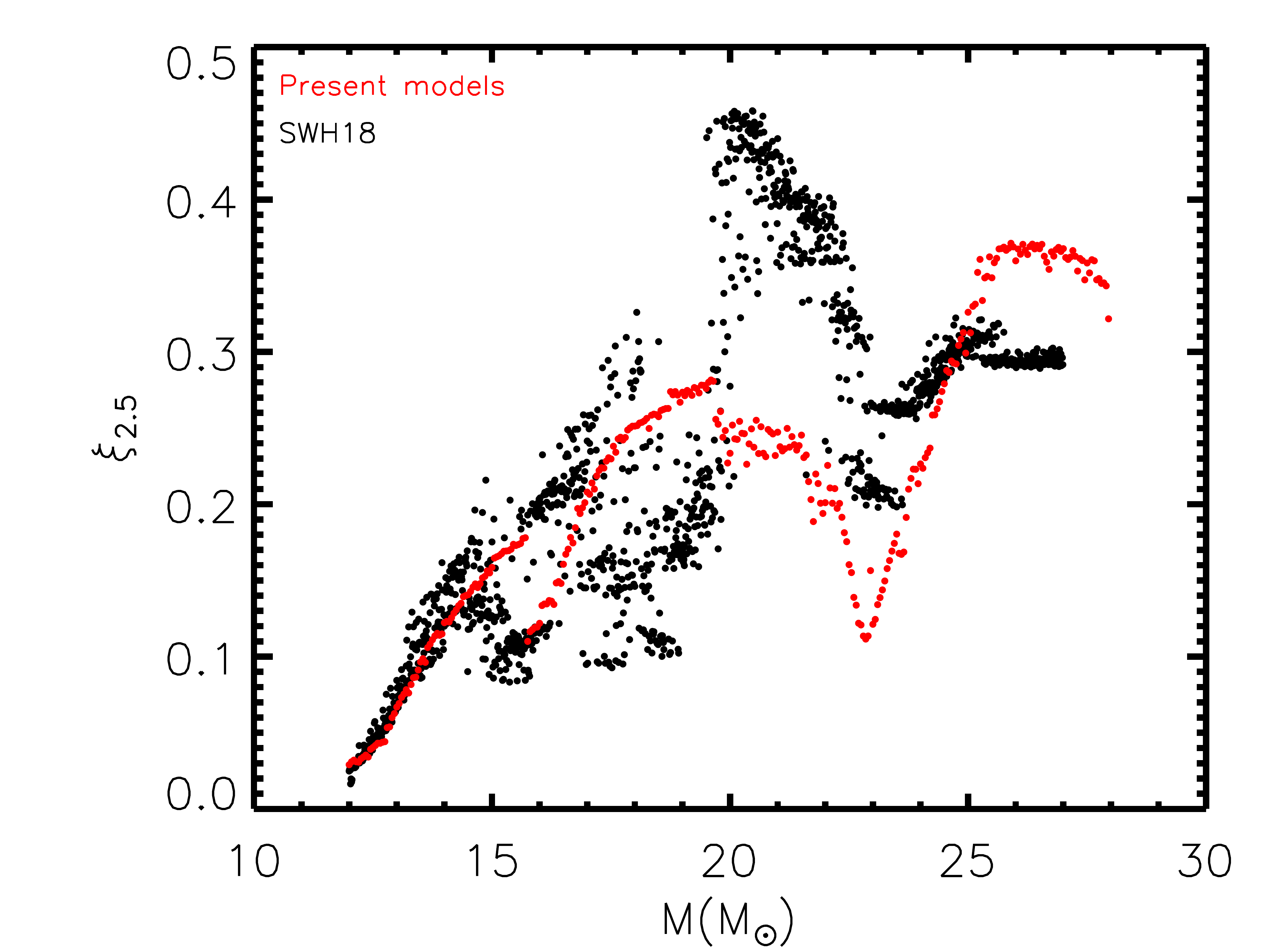}   {0.45\textwidth}{(e)}
          \fig{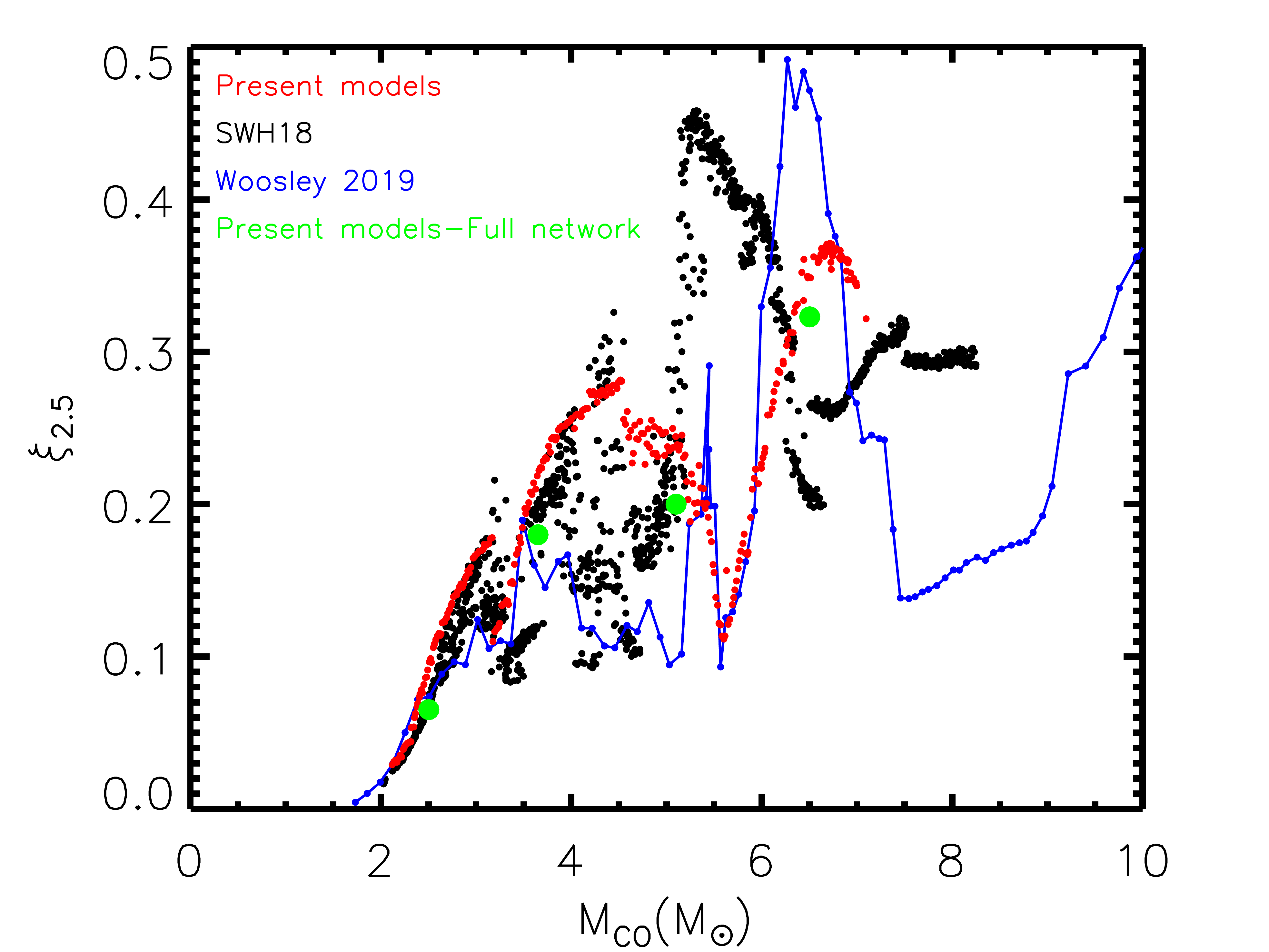}   {0.45\textwidth}{(f)}}
\caption{Comparison between some properties of our models and those of other similar computations available in the literature. {\bf Panel a} shows the fraction of Carbon left by the He burning in our models (red dots) and in those of SWH18 (blue dots). {\bf Panel b} shows the total mass, the He core mass, the CO core mass and the O burning shell mass. The respective color coding for our models is black, red, blue and green, while the corresponding coding for the SWH18 models is grey, magenta, cyan and dark green. {\bf Panel c} Same as panel {\it a} but with the addition of models published by \cite{wo19} and a few test models (see text). {\bf Panel d} Comparison between the $\rm M_{\rm CO}(M_{\rm He})$ relations of the present models and those of SWH18. {\bf Panel e} Comparison between the final compactness (computed for the mass coordinate M=2.5\msun) of our (red dots) and SWH18 (black dots) models. {\bf Panel f} Same as panel {\it e} but with the addition of the models by \cite{wo19} (blue dots and line) and a few test models (see text). \label{fig:conf1}}
\end{figure*}

Figure \ref{fig:conf1} shows the comparison between some key properties of our models and those published by SWH18. Panel {\it a} shows the run of the \nuk{C}{12} mass fraction left by the He burning with the initial mass, the red and blue dots referring to our and SWH18 models, respectively. It is evident that a quite large offset exists between the two sets of models. Since the evolution of a star in central He burning (and beyond) is largely controlled by its He core mass, and not the total mass, panel {\it c} in the same Figure shows the same comparison as a function of the He core mass. This panel is particularly robust because the conversion of C in O occurs towards the end of the He burning and since the final abundance of O scales directly with the central temperature (and hence with the He core mass), the final C/O ratio is largely fixed by the current value of the He core mass towards the end of the He burning and not by the previous history of the star. For example stars computed with or without mass loss are expected to lie basically on the same line in this kind of graph. The parameters that really control the final abundance of \nuk{C}{12} (for any fixed value of the He core mass) are the nuclear reaction rates of the $3\alpha$ and the \nuk{C}{12}($\alpha$,$\gamma$)\nuk{O}{16}, i.e. their  nuclear cross sections times the behavior of the convective core towards the end of the He burning  \citep{im01}.
The offset between the two sets of computations visible in panel {\it a} remains basically unaltered in panel {\it c}. Though both sets of models adopt the same (NACRE) nuclear cross section for the $3\alpha$, the nuclear cross section adopted for the \nuk{C}{12}($\alpha$,$\gamma$)\nuk{O}{16} is slightly different (we adopt \cite{kunzetal02} while SWH18 adopt 1.2 times \cite{bu96,bu97}, hereinafter BU961p2). In order to check the role played by the two different nuclear cross sections on the ashes of the He burning, we have recomputed three models (15, 20 and 27\msun) by adopting the BU961p2 nuclear cross section. The magenta dots in panel {\it c} refer to these test models: it is quite evident that at most one third of the offset may be due to the adoption of the two different nuclear cross sections. In our opinion the large offset is probably due to a substantial difference in the treatment of the border of the convective core in central He burning. A hint towards this explanation comes from panel {\it b} in Figure \ref{fig:conf1} where the final masses of the stars are shown as black (present models) and gray (SWH18) dots. Since most of the mass is lost during the H and He burning phases, the scatter present in the SWH18 models cannot depend on the advanced burning phases but on something occurring really in H/He burning. The authors discuss this point and state that this "noise" is due to an effect of semiconvection in central He burning that alters the surface properties of the stars and hence the mass loss rate. Note that such a "noise" leads to a quite large scatter in the final total mass for stars more massive than 17\msun~ or so and also to some scatter in the amount of \nuk{C}{12} left by the He burning. We cannot comment further this point, apart from noting that semiconvection in central He burning is very effective in low mass Horizontal Branch stars, and that it progressively becomes less important as the initial mass increases: above $\sim$10\msun~ or so, semiconvection should be negligible because of the progressive reduction of the dependence of the opacity on the C/He ratio \citep{ca85}. Instabilities that lead to the ingestion of fresh He in the core (usually referred to as Breathing Pulses, \citet{ca85}) may occur but are spurious phenomena, at least in the massive stars regime, that may be easily cured by a proper choice of the rezoning and the time step. Very recently \cite{wo19}, hereinafter W19, published a large set of models of bare He cores and his Figure 11 shows the amount of \nuk{C}{12} left by the He burning as a function of the He core mass. The set up of these computations is the same adopted by SWH18. Since, how we already discussed above, the final amount of C left by the He burning basically depends just on the He core mass during the latest phases of the He burning and not on the previous history of the star, it is meaningful to plot his results in panel {\it c}. The green dots represent the values obtained by W19 and are in excellent agreement with our three models computed with the same \nuk{C}{12}($\alpha$,$\gamma$)\nuk{O}{16} cross section adopted in the Kepler code.

In addition to the final total mass, panel {\it b} in Figure \ref{fig:conf1} shows also a comparison between the He core masses, the CO core masses and the O burning shell masses. The blue, red and green dots refer to our models while the cyan, magenta and dark green dots refer to the SWH18 models. Note that while the He and CO core masses of SWH18 show almost straight trends, our models bend slightly above 22\msun~ or so. Th reason is that stars more massive than 22\msun~ lose not only their H rich mantle but also part of their He core mass. Since the He burning depends on the He core mass, also the final CO core mass shows an analogous bend. Our models have He core masses systematically larger than those predicted SWH18 ones: this result is very probably connected to different choices for the determination of the border of the convective core in H/He burning. The actual size of the convective core (and convective shells) is still subject to serious uncertainties so that different choices are equally plausible. The run of the CO core masses versus the initial mass, vice versa, are in quite good agreement (apart from the more massive ones where the erosion of the He core due to mass loss induces the bending already discussed above), but this means that the He core mass - CO core mass relation is quite different between the two sets of models. To better highlight the differences between the two $\rm M_{\rm CO}(M_{\rm He})$ relations, panel {\it d} in the same Figure shows our relation as red dots and the SWH18 one as blue dots.

Since the fraction of \nuk{C}{12} left by the He burning and the $\rm M_{\rm CO}$ are the key parameters that drive all the advanced burning phase, the differences highlighted in panels {\it c} and {\it d} between the two sets of computations clearly indicate how difficult is to compare the final compactness predicted by the two sets of models. Therefore we simply show in panels {\it e} and {\it f} of Figure \ref{fig:conf1} a global comparison between the final $\rm \xi_{2.5}$ values: panel {\it e} shows the comparison as a function of the initial mass while panel {\it f} shows the same comparison as a function of the CO core mass. The red and black dots refer to our and SWH18 models, respectively. As expected the differences are quite large. Since the non monotonic average trend reflects the complex interplay among the various convective episodes, different combinations of CO core masses and \nuk{C}{12} abundances at the beginning of the advanced burning phases may easily lead to differences of the order of those shown in Figure \ref{fig:conf1}. However, it is worth noting that our results do not show any significant scatter around the main trend. This trend is very well defined and all the features shown by our models are well understood and discussed in Section \ref{sec:disc}.
A closer look to panel {\it f} in Figure \ref{fig:conf1} shows that the SWH18 and our models share some similarities. The compactness of the stars of lower mass, i.e. those having CO core masses up to, roughly, 3\msun~ is remarkably similar. The sharp discontinuity present in our models (largely discussed in the previous section) at $\rm M_{\rm CO}\sim3.3$\msun~ ($\rm M_{\rm ini}=15.75$\msun) is not present in the SWH18 models that, on the contrary, show a large scatter in this mass interval. However, note that a group of their models with low compactness clumps close to the position where our models show the discontinuity in the compactness $\rm \xi_{2.5}$. We will not attempt any further analysis because the large differences in the {\it initial conditions} at the beginning of the advanced burning phases prevent a reliable quantitative understanding of the different predictions. 
Since the models computed by W19 provides also their final compactness we show also their models in panel {\it f} of Figure \ref{fig:conf1}. These models are particularly useful because they present Carbon mass fraction intermediate between those obtained by us and those obtained by SWH18 (see above). The $\xi$ values of the W19 models are shown as blue dots connected by a blue line to increase visibility. There are obviously large differences because in any case the C mass fraction at the beginning of the advanced burning phases are significantly different but there are also striking similarities. In particular both the $\xi$ of the models in the low tail of the CO core mass (between, say, 1.5 and 3\msun) are remarkably similar (among all three sets of models) and the well shaped minimum around 5.5\msun. Also the maximum at 6/6.5\msun is quite similar even if the peak present in the W19 models is higher. The formation of a higher peak agrees with the general expectation that the lower the C mass fraction, the lower the efficiency of the C convective shell (the 3rd one) the more compact the star.

Before closing this section we want to mention a few tests we made to check the role played by the adoption of a small network instead of our usual very extended one \citep{lc18}. Though the amount of computer time necessary to run all these models with the full network is prohibitive for us, we computed 4 models (13, 18, 20 and 26\msun) with the full network. Note that our network (whichever is the size) is always fully coupled to the physical evolution and chemical mixing so that just one system of equation is solved each time step. In particular the system is formed by (4+number~of~isotopes)x(number~of~meshes), which means more that 1.5 millions of equations solved simultaneously for a network of 300 nuclear species and 5000 meshes. The cyan dots in panels {\it c} and {\it f} of Figure \ref{fig:conf1} show the C mass fraction left by the He burning and the final compactness of these four refined models. These tests show quite convincingly that the adoption of an extended, refined network does not change qualitatively the compactness obtained by means of a small network. 

\section{Conclusions} \label{sec:conc}
In this paper we presented a very fine grid (in mass) of models in the range 12 to 27.95\msun~ in order to look at the fine structure of the relation between initial mass and final compactness of the models. The evolution beyond the central He burning is bi-parametric because it depends on two parameters, the CO core mass and the fraction of C left by the He burning. In principle these two parameters are fully coupled (in non rotating stars) and not independent but, given the different prescriptions adopted by different groups in both managing convection and in the choice of the nuclear reaction rates, in practice there are in literature different pairings of CO core masses and fraction of C left by the He burning. 
Our models show that the compactness of a star, $\xi_{2.5}$, is strictly connected to the behavior, birth, growth, overlap and death, of the various C convective episodes. The relation $\rm \xi_{2.5}(M_{CO})$ is not a monotonic function of the CO core mass but shows features that are well understood and discussed. Moving from the low to the massive CO cores a first drastic change in the behavior of  $\xi_{2.5}$ occurs at $\rm M_{CO}\sim 3$\msun. The reason is that stars having CO core masses up to 3\msun~ or so must wait the disappearance of the second C convective shell before they can ignite Ne in the centre. CO core masses above 3\msun~, vice versa, are able to contract freely towards the Ne ignition independently on the ignition of the second C convective shell. As a consequence the second C convective shell ignites more violently that in the smaller masses causing the expansion of a large fraction of the mass above it. As the CO core mass increases, the strength of the second C convective shell progressively weakens (because of the inverse scaling of the fraction of C left by the He burning with the CO core mass) and the compactness of the star progressively increases again. However, as the CO core mass increases further, a second jump appears at a CO core mass of the order of 4.6\msun. This second jump is due to the progressive weakening of the efficiency of the second C convective shell that favors the contraction of the overlying mass and hence an early ignition of the third C convective shell. The net consequence is that the layers above this newly born C convective shell react by expanding and hence induce a reduction of the compactness $\xi_{2.5}$. As the CO core mass continues to increase the compactness starts raising again because also the strength of the third C convective shell progressively weakens as a consequence of the progressive lower C abundance left by the He burning.

Let us eventually stress again that all the features of the $\rm \xi_{2.5}(M_{CO})$ relation discussed above depend on the \nuk{C}{12}${\rm (M_{\rm CO})}$ relation, and therefore they can vary, even significantly, from one author to another. However, in spite of the complex interplay among the various C convective episodes that sculpt the dependence of the compactness of a star on the CO core mass, our models do not show any evidence of a significant scatter of the data: the relation is very tight and well defined.

\bigskip
\noindent
It is a pleasure to thank Stan Woosley and Tuguldur Sukhbold for having kindly provided their data in electronic form. This work has been partially supported by the italian grants "Premiale 2015 MITiC" (P.I. B. Garilli) and "Premiale 2015 FIGARO" (P.I. G. Gemme) and by the “ChETEC” COST Action (CA16117), supported by COST (European Cooperation in Science and Technology).

\clearpage
\startlongtable
\begin{deluxetable*}{cccccccccc}
\tabletypesize{\scriptsize}
\tablecolumns{10}
\tablecaption{Main data \label{tab:maindata}}
\tablehead{\\ $\rm M_{ini}$ & $\rm M_{fin}$ & $\rm M_{He}$ & $\rm M_{CO}$ & $\rm M_{Fe}$ & \nuk{C}{12} & $\rm \xi_{CO}$ & $\rm \xi_{2.5}$ \\
\msun&\msun&\msun&\msun&\msun&mass fraction&\msun/R($10^3$km)&\msun/R($10^3$km)}
\startdata
 12.00 & 10.7738 &  3.7111 &  2.1955 &  1.5001 &  0.3581 &  0.0721 &  0.0290 \\
 12.05 & 10.8002 &  3.7455 &  2.2161 &  1.5159 &  0.3619 &  0.0741 &  0.0309 \\
 12.10 & 10.8379 &  3.7689 &  2.2358 &  1.4475 &  0.3555 &  0.0709 &  0.0321 \\
 12.15 & 10.8922 &  3.7818 &  2.2412 &  1.5131 &  0.3611 &  0.0642 &  0.0310 \\
 12.20 & 10.9366 &  3.8001 &  2.2530 &  1.4402 &  0.3623 &  0.0595 &  0.0307 \\
 12.25 & 10.9767 &  3.8226 &  2.2685 &  1.4892 &  0.3625 &  0.0636 &  0.0330 \\
 12.30 & 11.0095 &  3.8510 &  2.2878 &  1.4419 &  0.3550 &  0.0615 &  0.0341 \\
 12.35 & 11.0641 &  3.8600 &  2.2942 &  1.4434 &  0.3608 &  0.0642 &  0.0354 \\
 12.40 & 11.1107 &  3.8818 &  2.3052 &  1.5078 &  0.3603 &  0.0572 &  0.0341 \\
 12.45 & 11.1369 &  3.9135 &  2.3272 &  1.4216 &  0.3590 &  0.0658 &  0.0392 \\
 12.50 & 11.1893 &  3.9292 &  2.3367 &  1.5247 &  0.3606 &  0.0663 &  0.0404 \\
 12.55 & 11.2314 &  3.9495 &  2.3510 &  1.4425 &  0.3590 &  0.0657 &  0.0417 \\
 12.60 & 11.2564 &  3.9849 &  2.3769 &  1.4767 &  0.3555 &  0.0621 &  0.0432 \\
 12.65 & 11.3186 &  3.9855 &  2.3800 &  1.5293 &  0.3528 &  0.0614 &  0.0431 \\
 12.70 & 11.3498 &  4.0169 &  2.3957 &  1.5354 &  0.3557 &  0.0591 &  0.0438 \\
 12.75 & 11.3915 &  4.0415 &  2.4128 &  1.5431 &  0.3496 &  0.0563 &  0.0442 \\
 12.80 & 11.4390 &  4.0571 &  2.4202 &  1.4482 &  0.3627 &  0.0702 &  0.0534 \\
 12.85 & 11.4431 &  4.0964 &  2.4522 &  1.5645 &  0.3541 &  0.0627 &  0.0538 \\
 12.90 & 11.4905 &  4.1130 &  2.4590 &  1.4487 &  0.3577 &  0.0693 &  0.0599 \\
 12.95 & 11.5369 &  4.1230 &  2.4682 &  1.5297 &  0.3583 &  0.0702 &  0.0626 \\
 13.00 & 11.5629 &  4.1553 &  2.4884 &  1.5251 &  0.3573 &  0.0698 &  0.0669 \\
 13.05 & 11.6241 &  4.1689 &  2.4968 &  1.5348 &  0.3591 &  0.0701 &  0.0692 \\
 13.10 & 11.6400 &  4.1973 &  2.5146 &  1.4719 &  0.3609 &  0.0701 &  0.0732 \\
 13.15 & 11.7076 &  4.2093 &  2.5230 &  1.5451 &  0.3619 &  0.0704 &  0.0754 \\
 13.20 & 11.7346 &  4.2317 &  2.5385 &  1.4707 &  0.3592 &  0.0697 &  0.0783 \\
 13.25 & 11.7848 &  4.2514 &  2.5511 &  1.4950 &  0.3616 &  0.0660 &  0.0761 \\
 13.30 & 11.8108 &  4.2762 &  2.5693 &  1.4732 &  0.3616 &  0.0672 &  0.0817 \\
 13.35 & 11.8397 &  4.3052 &  2.5874 &  1.5744 &  0.3612 &  0.0677 &  0.0862 \\
 13.40 & 11.8671 &  4.3288 &  2.6030 &  1.5654 &  0.3603 &  0.0660 &  0.0865 \\
 13.45 & 11.9208 &  4.3418 &  2.6140 &  1.5387 &  0.3609 &  0.0675 &  0.0912 \\
 13.50 & 11.9400 &  4.3692 &  2.6312 &  1.5373 &  0.3600 &  0.0684 &  0.0965 \\
 13.55 & 11.9701 &  4.3938 &  2.6495 &  1.5579 &  0.3591 &  0.0678 &  0.0987 \\
 13.60 & 12.0227 &  4.4074 &  2.6577 &  1.4453 &  0.3600 &  0.0660 &  0.0962 \\
 13.65 & 12.0381 &  4.4431 &  2.6807 &  1.5529 &  0.3593 &  0.0687 &  0.1059 \\
 13.70 & 12.0870 &  4.4601 &  2.6932 &  1.5468 &  0.3595 &  0.0690 &  0.1082 \\
 13.75 & 12.1171 &  4.4837 &  2.7112 &  1.4761 &  0.3597 &  0.0691 &  0.1111 \\
 13.80 & 12.1371 &  4.5071 &  2.7274 &  1.5452 &  0.3577 &  0.0693 &  0.1136 \\
 13.85 & 12.1516 &  4.5339 &  2.7462 &  1.5521 &  0.3577 &  0.0690 &  0.1153 \\
 13.90 & 12.1811 &  4.5563 &  2.7602 &  1.5505 &  0.3571 &  0.0681 &  0.1140 \\
 13.95 & 12.2555 &  4.5692 &  2.7701 &  1.5502 &  0.3574 &  0.0679 &  0.1149 \\
 14.00 & 12.2933 &  4.5918 &  2.7847 &  1.5518 &  0.3584 &  0.0701 &  0.1222 \\
 14.05 & 12.2610 &  4.6189 &  2.8056 &  1.5520 &  0.3577 &  0.0696 &  0.1231 \\
 14.10 & 12.3599 &  4.6307 &  2.8140 &  1.5569 &  0.3580 &  0.0691 &  0.1227 \\
 14.15 & 12.3450 &  4.6650 &  2.8368 &  1.5564 &  0.3568 &  0.0693 &  0.1255 \\
 14.20 & 12.3821 &  4.6858 &  2.8510 &  1.5552 &  0.3573 &  0.0698 &  0.1284 \\
 14.25 & 12.4220 &  4.7105 &  2.8695 &  1.5533 &  0.3566 &  0.0699 &  0.1307 \\
 14.30 & 12.4509 &  4.7419 &  2.8895 &  1.5547 &  0.3564 &  0.0703 &  0.1334 \\
 14.35 & 12.5129 &  4.7499 &  2.8996 &  1.6088 &  0.3565 &  0.0704 &  0.1350 \\
 14.40 & 12.5432 &  4.7788 &  2.9179 &  1.5608 &  0.3555 &  0.0714 &  0.1393 \\
 14.45 & 12.5827 &  4.8038 &  2.9330 &  1.5560 &  0.3557 &  0.0712 &  0.1401 \\
 14.50 & 12.6066 &  4.8362 &  2.9572 &  1.5641 &  0.3549 &  0.0707 &  0.1406 \\
 14.55 & 12.6676 &  4.8444 &  2.9657 &  1.5579 &  0.3549 &  0.0712 &  0.1427 \\
 14.60 & 12.6885 &  4.8782 &  2.9877 &  1.5674 &  0.3543 &  0.0715 &  0.1456 \\
 14.65 & 12.7279 &  4.8982 &  3.0030 &  1.5644 &  0.3536 &  0.0718 &  0.1477 \\
 14.70 & 12.7842 &  4.9189 &  3.0174 &  1.5684 &  0.3539 &  0.0708 &  0.1453 \\
 14.75 & 12.8162 &  4.9421 &  3.0333 &  1.5675 &  0.3537 &  0.0713 &  0.1479 \\
 14.80 & 12.8511 &  4.9703 &  3.0507 &  1.5644 &  0.3528 &  0.0718 &  0.1516 \\
 14.85 & 12.9187 &  4.9754 &  3.0578 &  1.6138 &  0.3542 &  0.0721 &  0.1527 \\
 14.90 & 12.9449 &  5.0070 &  3.0800 &  1.5620 &  0.3526 &  0.0727 &  0.1563 \\
 14.95 & 12.9874 &  5.0302 &  3.0944 &  1.5608 &  0.3534 &  0.0722 &  0.1549 \\
 15.00 & 13.0237 &  5.0528 &  3.1134 &  1.5090 &  0.3529 &  0.0727 &  0.1586 \\
 15.05 & 13.0493 &  5.0854 &  3.1391 &  1.5648 &  0.3519 &  0.0737 &  0.1647 \\
 15.10 & 13.0946 &  5.1075 &  3.1530 &  1.5668 &  0.3517 &  0.0738 &  0.1655 \\
 15.15 & 13.1587 &  5.1175 &  3.1625 &  1.5663 &  0.3514 &  0.0737 &  0.1662 \\
 15.20 & 13.1967 &  5.1482 &  3.1842 &  1.5666 &  0.3509 &  0.0737 &  0.1675 \\
 15.25 & 13.2153 &  5.1855 &  3.2080 &  1.5725 &  0.3506 &  0.0739 &  0.1691 \\
 15.30 & 13.2661 &  5.1997 &  3.2227 &  1.5744 &  0.3503 &  0.0738 &  0.1693 \\
 15.35 & 13.3142 &  5.2175 &  3.2349 &  1.5746 &  0.3500 &  0.0737 &  0.1695 \\
 15.40 & 13.3432 &  5.2427 &  3.2515 &  1.5661 &  0.3501 &  0.0739 &  0.1704 \\
 15.45 & 13.3921 &  5.2572 &  3.2658 &  1.5698 &  0.3503 &  0.0744 &  0.1737 \\
 15.50 & 13.4323 &  5.2822 &  3.2817 &  1.5700 &  0.3491 &  0.0742 &  0.1729 \\
 15.55 & 13.4618 &  5.3036 &  3.2965 &  1.5722 &  0.3495 &  0.0742 &  0.1735 \\
 15.60 & 13.4981 &  5.3307 &  3.3176 &  1.5731 &  0.3488 &  0.0740 &  0.1742 \\
 15.65 & 13.5139 &  5.3614 &  3.3386 &  1.6299 &  0.3481 &  0.0746 &  0.1775 \\
 15.70 & 13.5610 &  5.3819 &  3.3546 &  1.5712 &  0.3484 &  0.0747 &  0.1780 \\
 15.75 & 13.6051 &  5.4033 &  3.3878 &  1.4803 &  0.3478 &  0.0528 &  0.1100 \\
 15.80 & 13.6237 &  5.4281 &  3.4054 &  1.5413 &  0.3469 &  0.0551 &  0.1164 \\
 15.85 & 13.6532 &  5.4545 &  3.4236 &  1.4981 &  0.3467 &  0.0555 &  0.1178 \\
 15.90 & 13.7030 &  5.4792 &  3.4400 &  1.4754 &  0.3468 &  0.0559 &  0.1191 \\
 15.95 & 13.7520 &  5.4919 &  3.4531 &  1.5361 &  0.3466 &  0.0558 &  0.1194 \\
 16.00 & 13.7959 &  5.5046 &  3.4633 &  1.5334 &  0.3466 &  0.0566 &  0.1219 \\
 16.05 & 13.8168 &  5.5357 &  3.4893 &  1.5073 &  0.3457 &  0.0608 &  0.1335 \\
 16.10 & 13.8443 &  5.5645 &  3.5063 &  1.5450 &  0.3455 &  0.0611 &  0.1345 \\
 16.15 & 13.8764 &  5.5889 &  3.5248 &  1.5454 &  0.3454 &  0.0609 &  0.1346 \\
 16.20 & 13.9088 &  5.6112 &  3.5403 &  1.5429 &  0.3451 &  0.0616 &  0.1368 \\
 16.25 &  5.6905 &  5.6233 &  3.5579 &  1.5463 &  0.3460 &  0.0612 &  0.1364 \\
 16.30 &  5.7003 &  5.6342 &  3.5657 &  1.5175 &  0.3467 &  0.0605 &  0.1342 \\
 16.35 &  5.7342 &  5.6641 &  3.5915 &  1.5472 &  0.3450 &  0.0653 &  0.1484 \\
 16.40 &  5.7583 &  5.6894 &  3.6080 &  1.5518 &  0.3450 &  0.0656 &  0.1492 \\
 16.45 &  5.7740 &  5.7035 &  3.6210 &  1.5475 &  0.3451 &  0.0650 &  0.1480 \\
 16.50 &  5.7927 &  5.7225 &  3.6369 &  1.5544 &  0.3442 &  0.0696 &  0.1606 \\
 16.55 &  5.8206 &  5.7506 &  3.6584 &  1.5557 &  0.3442 &  0.0717 &  0.1672 \\
 16.60 &  5.8479 &  5.7785 &  3.6773 &  1.4769 &  0.3436 &  0.0726 &  0.1705 \\
 16.65 &  5.8769 &  5.8029 &  3.6974 &  1.5651 &  0.3429 &  0.0744 &  0.1782 \\
 16.70 &  5.8854 &  5.8181 &  3.7072 &  1.5712 &  0.3433 &  0.0735 &  0.1746 \\
 16.75 &  5.9120 &  5.8428 &  3.7281 &  1.5682 &  0.3423 &  0.0761 &  0.1846 \\
 16.80 &  5.9505 &  5.8814 &  3.7552 &  1.5729 &  0.3413 &  0.0791 &  0.1972 \\
 16.85 &  5.9609 &  5.8907 &  3.7637 &  1.5715 &  0.3418 &  0.0784 &  0.1939 \\
 16.90 &  5.9849 &  5.9103 &  3.7814 &  1.5672 &  0.3415 &  0.0794 &  0.1977 \\
 16.95 &  6.0056 &  5.9374 &  3.7994 &  1.5695 &  0.3412 &  0.0801 &  0.2010 \\
 17.00 &  6.0370 &  5.9658 &  3.8220 &  1.5671 &  0.3400 &  0.0815 &  0.2080 \\
 17.05 &  6.0519 &  5.9788 &  3.8358 &  1.5731 &  0.3401 &  0.0811 &  0.2065 \\
 17.10 &  6.0807 &  6.0104 &  3.8575 &  1.5706 &  0.3395 &  0.0825 &  0.2140 \\
 17.15 &  6.1098 &  6.0376 &  3.8779 &  1.5723 &  0.3391 &  0.0815 &  0.2099 \\
 17.20 &  6.1320 &  6.0585 &  3.8979 &  1.5721 &  0.3388 &  0.0833 &  0.2186 \\
 17.25 &  6.1554 &  6.0811 &  3.9150 &  1.5662 &  0.3378 &  0.0840 &  0.2222 \\
 17.30 &  6.1770 &  6.1025 &  3.9316 &  1.5578 &  0.3375 &  0.0841 &  0.2240 \\
 17.35 &  6.1991 &  6.1236 &  3.9500 &  1.5335 &  0.3373 &  0.0837 &  0.2237 \\
 17.40 &  6.2204 &  6.1470 &  3.9667 &  1.5240 &  0.3369 &  0.0849 &  0.2283 \\
 17.45 &  6.2556 &  6.1786 &  3.9906 &  1.5298 &  0.3358 &  0.0856 &  0.2305 \\
 17.50 &  6.2691 &  6.1982 &  4.0021 &  1.6336 &  0.3357 &  0.0856 &  0.2298 \\
 17.55 &  6.2987 &  6.2278 &  4.0277 &  1.5055 &  0.3355 &  0.0865 &  0.2381 \\
 17.60 &  6.3215 &  6.2454 &  4.0468 &  1.5088 &  0.3347 &  0.0854 &  0.2341 \\
 17.65 &  6.3434 &  6.2683 &  4.0627 &  1.5532 &  0.3343 &  0.0873 &  0.2429 \\
 17.70 &  6.3660 &  6.2903 &  4.0810 &  1.5575 &  0.3343 &  0.0877 &  0.2440 \\
 17.75 &  6.3986 &  6.3234 &  4.1050 &  1.4925 &  0.3336 &  0.0872 &  0.2422 \\
 17.80 &  6.4105 &  6.3379 &  4.1176 &  1.5727 &  0.3334 &  0.0875 &  0.2438 \\
 17.85 &  6.4384 &  6.3663 &  4.1378 &  1.5825 &  0.3327 &  0.0885 &  0.2487 \\
 17.90 &  6.4603 &  6.3824 &  4.1554 &  1.5880 &  0.3325 &  0.0889 &  0.2501 \\
 17.95 &  6.4861 &  6.4087 &  4.1756 &  1.5989 &  0.3317 &  0.0891 &  0.2512 \\
 18.00 &  6.5151 &  6.4367 &  4.1968 &  1.6053 &  0.3312 &  0.0893 &  0.2513 \\
 18.05 &  6.5356 &  6.4568 &  4.2154 &  1.6177 &  0.3310 &  0.0893 &  0.2516 \\
 18.10 &  6.5629 &  6.4871 &  4.2330 &  1.6293 &  0.3303 &  0.0900 &  0.2533 \\
 18.15 &  6.5754 &  6.4969 &  4.2466 &  1.6237 &  0.3306 &  0.0900 &  0.2538 \\
 18.20 &  6.6068 &  6.5318 &  4.2708 &  1.6433 &  0.3298 &  0.0902 &  0.2547 \\
 18.25 &  6.6280 &  6.5490 &  4.2890 &  1.6433 &  0.3298 &  0.0906 &  0.2563 \\
 18.30 &  6.6695 &  6.5901 &  4.3189 &  1.5726 &  0.3286 &  0.0886 &  0.2497 \\
 18.35 &  6.6853 &  6.6082 &  4.3344 &  1.6492 &  0.3285 &  0.0910 &  0.2585 \\
 18.40 &  6.7068 &  6.6263 &  4.3507 &  1.6496 &  0.3280 &  0.0906 &  0.2591 \\
 18.45 &  6.7316 &  6.6522 &  4.3694 &  1.5979 &  0.3274 &  0.0906 &  0.2588 \\
 18.50 &  6.7608 &  6.6840 &  4.3948 &  1.6079 &  0.3266 &  0.0901 &  0.2575 \\
 18.55 &  6.7759 &  6.6983 &  4.4071 &  1.6065 &  0.3266 &  0.0910 &  0.2616 \\
 18.60 &  6.8029 &  6.7264 &  4.4294 &  1.6108 &  0.3256 &  0.0911 &  0.2622 \\
 18.65 &  6.8236 &  6.7442 &  4.4480 &  1.6075 &  0.3251 &  0.0909 &  0.2628 \\
 18.70 &  6.8497 &  6.7744 &  4.4694 &  1.5964 &  0.3243 &  0.0907 &  0.2629 \\
 18.75 &  6.8742 &  6.7923 &  4.4880 &  1.6113 &  0.3234 &  0.0927 &  0.2739 \\
 18.80 &  6.8996 &  6.8205 &  4.5086 &  1.6253 &  0.3231 &  0.0923 &  0.2718 \\
 18.85 &  6.9229 &  6.8401 &  4.5292 &  1.5869 &  0.3223 &  0.0926 &  0.2739 \\
 18.90 &  6.9546 &  6.8755 &  4.5566 &  1.5979 &  0.3212 &  0.0916 &  0.2717 \\
 18.95 &  6.9711 &  6.8902 &  4.5711 &  1.6665 &  0.3207 &  0.0906 &  0.2669 \\
 19.00 &  6.9973 &  6.9177 &  4.5943 &  1.6416 &  0.3198 &  0.0920 &  0.2735 \\
 19.05 &  7.0148 &  6.9307 &  4.6083 &  1.5234 &  0.3196 &  0.0916 &  0.2713 \\
 19.10 &  7.0400 &  6.9633 &  4.6281 &  1.6532 &  0.3193 &  0.0923 &  0.2747 \\
 19.15 &  7.0676 &  6.9891 &  4.6519 &  1.6634 &  0.3183 &  0.0916 &  0.2737 \\
 19.20 &  7.0904 &  7.0066 &  4.6690 &  1.6638 &  0.3177 &  0.0917 &  0.2715 \\
 19.25 &  7.1159 &  7.0388 &  4.6948 &  1.6632 &  0.3166 &  0.0925 &  0.2766 \\
 19.30 &  7.1400 &  7.0608 &  4.7119 &  1.6318 &  0.3167 &  0.0921 &  0.2759 \\
 19.35 &  7.1619 &  7.0818 &  4.7339 &  1.6260 &  0.3156 &  0.0909 &  0.2731 \\
 19.40 &  7.1831 &  7.1034 &  4.7537 &  1.6442 &  0.3152 &  0.0919 &  0.2782 \\
 19.45 &  7.2053 &  7.1221 &  4.7664 &  1.6624 &  0.3151 &  0.0919 &  0.2780 \\
 19.50 &  7.2334 &  7.1535 &  4.7909 &  1.6286 &  0.3142 &  0.0915 &  0.2760 \\
 19.55 &  7.2533 &  7.1712 &  4.8108 &  1.6539 &  0.3137 &  0.0918 &  0.2801 \\
 19.60 &  7.2807 &  7.1979 &  4.8307 &  1.6581 &  0.3131 &  0.0919 &  0.2815 \\
 19.65 &  7.2977 &  7.2142 &  4.8482 &  1.6644 &  0.3134 &  0.0918 &  0.2807 \\
 19.70 &  7.3215 &  7.2383 &  4.8657 &  1.6686 &  0.3129 &  0.0892 &  0.2557 \\
 19.75 &  7.3446 &  7.2640 &  4.8896 &  1.6509 &  0.3119 &  0.0886 &  0.2525 \\
 19.80 &  7.3677 &  7.2854 &  4.9048 &  1.6541 &  0.3121 &  0.0900 &  0.2609 \\
 19.85 &  7.3893 &  7.3074 &  4.9259 &  1.6485 &  0.3117 &  0.0870 &  0.2440 \\
 19.90 &  7.4120 &  7.3287 &  4.9497 &  1.6548 &  0.3113 &  0.0869 &  0.2483 \\
 19.95 &  7.4367 &  7.3538 &  4.9661 &  1.6590 &  0.3106 &  0.0843 &  0.2271 \\
 20.00 &  7.4517 &  7.3692 &  4.9806 &  1.6356 &  0.3107 &  0.0853 &  0.2334 \\
 20.05 &  7.4772 &  7.3925 &  4.9960 &  1.6492 &  0.3105 &  0.0870 &  0.2520 \\
 20.10 &  7.5040 &  7.4208 &  5.0202 &  1.6284 &  0.3100 &  0.0860 &  0.2429 \\
 20.15 &  7.5268 &  7.4418 &  5.0375 &  1.6507 &  0.3099 &  0.0854 &  0.2425 \\
 20.20 &  7.5455 &  7.4600 &  5.0549 &  1.6481 &  0.3096 &  0.0862 &  0.2544 \\
 20.25 &  7.5748 &  7.4862 &  5.0800 &  1.6618 &  0.3089 &  0.0854 &  0.2466 \\
 20.30 &  7.5965 &  7.5105 &  5.0993 &  1.6166 &  0.3090 &  0.0854 &  0.2463 \\
 20.35 &  7.6223 &  7.5352 &  5.1162 &  1.5614 &  0.3080 &  0.0826 &  0.2262 \\
 20.40 &  7.6432 &  7.5544 &  5.1367 &  1.5789 &  0.3085 &  0.0843 &  0.2396 \\
 20.45 &  7.6660 &  7.5781 &  5.1538 &  1.5710 &  0.3083 &  0.0851 &  0.2495 \\
 20.50 &  7.6874 &  7.5952 &  5.1714 &  1.5894 &  0.3076 &  0.0839 &  0.2374 \\
 20.55 &  7.7091 &  7.6161 &  5.1909 &  1.6619 &  0.3085 &  0.0853 &  0.2551 \\
 20.60 &  7.7348 &  7.6448 &  5.2088 &  1.5892 &  0.3076 &  0.0829 &  0.2333 \\
 20.65 &  7.7521 &  7.6635 &  5.2238 &  1.5994 &  0.3086 &  0.0849 &  0.2509 \\
 20.70 &  7.7754 &  7.6821 &  5.2435 &  1.5912 &  0.3082 &  0.0826 &  0.2335 \\
 20.75 &  7.7940 &  7.7045 &  5.2611 &  1.5872 &  0.3071 &  0.0817 &  0.2312 \\
 20.80 &  7.8198 &  7.7302 &  5.2783 &  1.5953 &  0.3080 &  0.0839 &  0.2491 \\
 20.85 &  7.8486 &  7.7566 &  5.2998 &  1.5914 &  0.3075 &  0.0828 &  0.2474 \\
 20.90 &  7.8712 &  7.7807 &  5.3146 &  1.5897 &  0.3075 &  0.0828 &  0.2462 \\
 20.95 &  7.8958 &  7.8057 &  5.3377 &  1.5855 &  0.3070 &  0.0809 &  0.2320 \\
 21.00 &  7.9181 &  7.8233 &  5.3505 &  1.5880 &  0.3071 &  0.0825 &  0.2473 \\
 21.05 &  7.9377 &  7.8527 &  5.3729 &  1.5825 &  0.3071 &  0.0814 &  0.2377 \\
 21.10 &  7.9553 &  7.8736 &  5.3926 &  1.5802 &  0.3065 &  0.0802 &  0.2350 \\
 21.15 &  7.9817 &  7.9012 &  5.4098 &  1.5496 &  0.3068 &  0.0807 &  0.2364 \\
 21.20 &  7.9931 &  7.9145 &  5.4320 &  1.6065 &  0.3064 &  0.0817 &  0.2498 \\
 21.25 &  8.0148 &  7.9349 &  5.4455 &  1.5018 &  0.3059 &  0.0802 &  0.2375 \\
 21.30 &  8.0121 &  7.9330 &  5.4680 &  1.5860 &  0.3067 &  0.0804 &  0.2444 \\
 21.35 &  8.0430 &  7.9632 &  5.4808 &  1.5709 &  0.3062 &  0.0796 &  0.2391 \\
 21.40 &  8.0510 &  7.9689 &  5.5047 &  1.5692 &  0.3065 &  0.0794 &  0.2355 \\
 21.45 &  8.0625 &  7.9809 &  5.5169 &  1.5668 &  0.3061 &  0.0793 &  0.2385 \\
 21.50 &  8.0932 &  8.0124 &  5.5364 &  1.5916 &  0.3055 &  0.0794 &  0.2454 \\
 21.55 &  8.1049 &  8.0209 &  5.5591 &  1.5675 &  0.3052 &  0.0774 &  0.2306 \\
 21.60 &  8.1246 &  8.0424 &  5.5769 &  1.5654 &  0.3053 &  0.0775 &  0.2323 \\
 21.65 &  8.1347 &  8.0523 &  5.5972 &  1.6268 &  0.3050 &  0.0758 &  0.2149 \\
 21.70 &  8.1508 &  8.0672 &  5.6101 &  1.6473 &  0.3053 &  0.0745 &  0.2031 \\
 21.75 &  8.1713 &  8.0833 &  5.6366 &  1.6530 &  0.3045 &  0.0715 &  0.1886 \\
 21.80 &  8.1928 &  8.1040 &  5.6496 &  1.5751 &  0.3049 &  0.0751 &  0.2198 \\
 21.85 &  8.2061 &  8.1190 &  5.6687 &  1.5501 &  0.3048 &  0.0739 &  0.2136 \\
 21.90 &  8.2159 &  8.1338 &  5.6910 &  1.5337 &  0.3042 &  0.0716 &  0.2007 \\
 21.95 &  8.2330 &  8.1475 &  5.7099 &  1.6295 &  0.3039 &  0.0716 &  0.1940 \\
 22.00 &  8.2529 &  8.1695 &  5.7256 &  1.5476 &  0.3038 &  0.0716 &  0.2010 \\
 22.05 &  8.2508 &  8.1678 &  5.7394 &  1.6593 &  0.3040 &  0.0760 &  0.2255 \\
 22.10 &  8.2711 &  8.1845 &  5.7587 &  1.6669 &  0.3037 &  0.0739 &  0.2107 \\
 22.15 &  8.2919 &  8.2083 &  5.7811 &  1.5984 &  0.3034 &  0.0723 &  0.2007 \\
 22.20 &  8.3036 &  8.2157 &  5.8017 &  1.6363 &  0.3030 &  0.0730 &  0.2103 \\
 22.25 &  8.3000 &  8.1722 &  5.8151 &  1.5734 &  0.3029 &  0.0707 &  0.1973 \\
 22.30 &  8.3208 &  8.1932 &  5.8336 &  1.5452 &  0.3026 &  0.0710 &  0.2005 \\
 22.35 &  8.3269 &  8.2083 &  5.8530 &  1.5396 &  0.3025 &  0.0695 &  0.1915 \\
 22.40 &  8.3428 &  8.2389 &  5.8721 &  1.5300 &  0.3023 &  0.0679 &  0.1813 \\
 22.45 &  8.3532 &  8.2612 &  5.8932 &  1.5933 &  0.3024 &  0.0667 &  0.1754 \\
 22.50 &  8.3688 &  8.2801 &  5.9165 &  1.5621 &  0.3018 &  0.0638 &  0.1603 \\
 22.55 &  8.3765 &  8.2922 &  5.9297 &  1.6188 &  0.3020 &  0.0626 &  0.1552 \\
 22.60 &  8.4000 &  8.3152 &  5.9518 &  1.5399 &  0.3015 &  0.0596 &  0.1389 \\
 22.65 &  8.4054 &  8.3210 &  5.9727 &  1.4915 &  0.3013 &  0.0585 &  0.1338 \\
 22.70 &  8.4128 &  8.3283 &  5.9907 &  1.5333 &  0.3011 &  0.0562 &  0.1219 \\
 22.75 &  8.4181 &  8.3337 &  6.0081 &  1.5097 &  0.3009 &  0.0558 &  0.1208 \\
 22.80 &  8.4498 &  8.3649 &  6.0323 &  1.4487 &  0.3004 &  0.0565 &  0.1135 \\
 22.85 &  8.4566 &  8.3716 &  6.0501 &  1.2922 &  0.3002 &  0.0553 &  0.1115 \\
 22.90 &  8.4663 &  8.3808 &  6.0630 &  1.3110 &  0.3001 &  0.0566 &  0.1135 \\
 22.95 &  8.4783 &  8.3931 &  6.0834 &  1.5263 &  0.2999 &  0.0728 &  0.1565 \\
 23.00 &  8.4979 &  8.4128 &  6.1002 &  1.5113 &  0.2994 &  0.0567 &  0.1212 \\
 23.05 &  8.4987 &  8.4137 &  6.1187 &  1.4816 &  0.2992 &  0.0573 &  0.1244 \\
 23.10 &  8.5246 &  8.4388 &  6.1425 &  1.5036 &  0.2988 &  0.0595 &  0.1341 \\
 23.15 &  8.5334 &  8.4478 &  6.1571 &  1.4911 &  0.2985 &  0.0606 &  0.1387 \\
 23.20 &  8.5438 &  8.4580 &  6.1719 &  1.4932 &  0.2984 &  0.0617 &  0.1438 \\
 23.25 &  8.5556 &  8.4700 &  6.1903 &  1.5006 &  0.2982 &  0.0630 &  0.1496 \\
 23.30 &  8.5638 &  8.4780 &  6.2129 &  1.5662 &  0.2978 &  0.0647 &  0.1578 \\
 23.35 &  8.5844 &  8.4982 &  6.2302 &  1.5409 &  0.2975 &  0.0657 &  0.1631 \\
 23.40 &  8.5930 &  8.5068 &  6.2488 &  1.5112 &  0.2973 &  0.0669 &  0.1692 \\
 23.45 &  8.6165 &  8.5302 &  6.2675 &  1.6295 &  0.2969 &  0.0679 &  0.1743 \\
 23.50 &  8.6113 &  8.5252 &  6.2862 &  1.6301 &  0.2967 &  0.0691 &  0.1804 \\
 23.55 &  8.6034 &  8.5169 &  6.3049 &  1.6085 &  0.2964 &  0.0671 &  0.1678 \\
 23.60 &  8.6504 &  8.5636 &  6.3237 &  1.6516 &  0.2961 &  0.0683 &  0.1671 \\
 23.65 &  8.6341 &  8.5475 &  6.3317 &  1.6599 &  0.2962 &  0.0686 &  0.1686 \\
 23.70 &  8.6753 &  8.5883 &  6.3612 &  1.5868 &  0.2954 &  0.0722 &  0.1913 \\
 23.75 &  8.6830 &  8.5961 &  6.3854 &  1.6358 &  0.2950 &  0.0746 &  0.2099 \\
 23.80 &  8.7057 &  8.6183 &  6.4043 &  1.6441 &  0.2949 &  0.0755 &  0.2170 \\
 23.85 &  8.7326 &  8.6449 &  6.4286 &  1.6456 &  0.2943 &  0.0764 &  0.2230 \\
 23.90 &  8.6843 &  8.5970 &  6.4312 &  1.5816 &  0.2946 &  0.0763 &  0.2229 \\
 23.95 &  8.7380 &  8.6504 &  6.4609 &  1.6397 &  0.2938 &  0.0763 &  0.2135 \\
 24.00 &  8.7830 &  8.6951 &  6.4908 &  1.6657 &  0.2934 &  0.0783 &  0.2265 \\
 24.05 &  8.7635 &  8.6758 &  6.4934 &  1.6721 &  0.2936 &  0.0779 &  0.2236 \\
 24.10 &  8.7767 &  8.6886 &  6.5178 &  1.6717 &  0.2930 &  0.0789 &  0.2306 \\
 24.15 &  8.7824 &  8.6945 &  6.5334 &  1.6702 &  0.2931 &  0.0795 &  0.2338 \\
 24.20 &  8.7776 &  8.6896 &  6.5469 &  1.6757 &  0.2930 &  0.0799 &  0.2368 \\
 24.25 &  8.8099 &  8.7213 &  6.5749 &  1.6751 &  0.2924 &  0.0820 &  0.2586 \\
 24.30 &  8.8292 &  8.7405 &  6.5940 &  1.6707 &  0.2921 &  0.0824 &  0.2587 \\
 24.35 &  8.8681 &  8.7793 &  6.6234 &  1.6542 &  0.2914 &  0.0847 &  0.2626 \\
 24.40 &  8.8679 &  8.7792 &  6.6405 &  1.6546 &  0.2912 &  0.0856 &  0.2673 \\
 24.45 &  8.8679 &  8.7788 &  6.6486 &  1.6810 &  0.2913 &  0.0860 &  0.2740 \\
 24.50 &  8.8930 &  8.8038 &  6.6761 &  1.6814 &  0.2908 &  0.0879 &  0.2791 \\
 24.55 &  8.9085 &  8.8192 &  6.6953 &  1.6808 &  0.2906 &  0.0892 &  0.2878 \\
 24.60 &  8.9284 &  8.8390 &  6.7201 &  1.6742 &  0.2901 &  0.0900 &  0.2865 \\
 24.65 &  8.9750 &  8.8852 &  6.7449 &  1.6917 &  0.2895 &  0.0925 &  0.2940 \\
 24.70 &  8.9536 &  8.8640 &  6.7481 &  1.6810 &  0.2897 &  0.0925 &  0.2926 \\
 24.75 &  8.9391 &  8.8493 &  6.7554 &  1.6895 &  0.2899 &  0.0923 &  0.2921 \\
 24.80 &  8.9899 &  8.9000 &  6.7860 &  1.6919 &  0.2892 &  0.0953 &  0.3042 \\
 24.85 &  8.9974 &  8.9073 &  6.8053 &  1.6831 &  0.2891 &  0.0960 &  0.3084 \\
 24.90 &  9.0062 &  8.9161 &  6.8296 &  1.6732 &  0.2886 &  0.0972 &  0.3127 \\
 24.95 &  9.0342 &  8.9438 &  6.8507 &  1.6776 &  0.2883 &  0.0946 &  0.2992 \\
 25.00 &  9.0540 &  8.9629 &  6.8698 &  1.7065 &  0.2879 &  0.1009 &  0.3260 \\
 25.05 &  9.0267 &  8.9363 &  6.8719 &  1.7212 &  0.2885 &  0.0978 &  0.3125 \\
 25.10 &  9.0462 &  8.9554 &  6.8965 &  1.7021 &  0.2881 &  0.1011 &  0.3299 \\
 25.15 &  9.0626 &  8.9718 &  6.9190 &  1.7075 &  0.2877 &  0.1017 &  0.3313 \\
 25.20 &  9.1074 &  9.0159 &  6.9482 &  1.7181 &  0.2873 &  0.1067 &  0.3522 \\
 25.25 &  9.1257 &  9.0343 &  6.9716 &  1.7215 &  0.2869 &  0.1096 &  0.3607 \\
 25.30 &  9.1185 &  9.0269 &  6.9845 &  1.7043 &  0.2870 &  0.1013 &  0.3337 \\
 25.35 &  9.1762 &  9.0843 &  7.0199 &  1.7035 &  0.2863 &  0.1039 &  0.3486 \\
 25.40 &  9.1603 &  9.0684 &  7.0265 &  1.7141 &  0.2865 &  0.1035 &  0.3496 \\
 25.45 &  9.2480 &  9.1551 &  7.0736 &  1.7185 &  0.2852 &  0.1099 &  0.3624 \\
 25.50 &  9.1824 &  9.0903 &  7.0687 &  1.7136 &  0.2858 &  0.1041 &  0.3486 \\
 25.55 &  9.2957 &  9.2025 &  7.1195 &  1.7127 &  0.2844 &  0.1110 &  0.3585 \\
 25.60 &  9.2636 &  9.1707 &  7.1240 &  1.7328 &  0.2849 &  0.1118 &  0.3614 \\
 25.65 &  9.3774 &  9.2835 &  7.1885 &  1.7257 &  0.2827 &  0.1176 &  0.3676 \\
 25.70 &  9.3924 &  9.2981 &  7.2041 &  1.7312 &  0.2827 &  0.1190 &  0.3678 \\
 25.75 &  9.3770 &  9.2829 &  7.2160 &  1.7283 &  0.2827 &  0.1177 &  0.3687 \\
 25.80 &  9.4043 &  9.3100 &  7.2419 &  1.7316 &  0.2824 &  0.1189 &  0.3667 \\
 25.85 &  9.2841 &  9.1911 &  7.1999 &  1.7286 &  0.2847 &  0.1137 &  0.3680 \\
 25.90 &  9.4188 &  9.3244 &  7.2683 &  1.7299 &  0.2825 &  0.1202 &  0.3713 \\
 25.95 &  9.4012 &  9.3070 &  7.2763 &  1.7290 &  0.2828 &  0.1187 &  0.3689 \\
 26.00 &  9.2422 &  9.1495 &  6.6617 &  1.7227 &  0.2854 &  0.1292 &  0.3598 \\
 26.05 &  9.2835 &  9.1904 &  6.6741 &  1.7314 &  0.2852 &  0.1377 &  0.3675 \\
 26.10 &  9.2856 &  9.1926 &  6.7128 &  1.7198 &  0.2852 &  0.1343 &  0.3641 \\
 26.15 &  9.4289 &  9.3345 &  6.8422 &  1.7369 &  0.2826 &  0.1421 &  0.3706 \\
 26.20 &  9.3158 &  9.2222 &  6.7702 &  1.7296 &  0.2849 &  0.1348 &  0.3667 \\
 26.25 &  9.3252 &  9.2319 &  6.7710 &  1.7243 &  0.2849 &  0.1351 &  0.3639 \\
 26.30 &  9.3541 &  9.2603 &  6.8531 &  1.7329 &  0.2846 &  0.1356 &  0.3691 \\
 26.35 &  9.3620 &  9.2684 &  6.8571 &  1.7352 &  0.2847 &  0.1360 &  0.3707 \\
 26.40 &  9.3527 &  9.2592 &  6.9091 &  1.7514 &  0.2848 &  0.1298 &  0.3676 \\
 26.45 &  9.3772 &  9.2832 &  6.9335 &  1.7342 &  0.2844 &  0.1313 &  0.3699 \\
 26.50 &  9.3846 &  9.2903 &  6.9353 &  1.7304 &  0.2849 &  0.1319 &  0.3677 \\
 26.55 &  9.3992 &  9.3050 &  6.9544 &  1.7347 &  0.2846 &  0.1328 &  0.3706 \\
 26.60 &  9.3645 &  9.2708 &  6.9518 &  1.7420 &  0.2857 &  0.1315 &  0.3628 \\
 26.65 &  9.4422 &  9.3474 &  7.0093 &  1.7190 &  0.2837 &  0.1364 &  0.3589 \\
 26.70 &  9.4482 &  9.3536 &  7.0179 &  1.7324 &  0.2839 &  0.1353 &  0.3542 \\
 26.75 &  9.4433 &  9.3485 &  7.0252 &  1.7316 &  0.2841 &  0.1350 &  0.3656 \\
 26.80 &  9.4718 &  9.3770 &  7.0381 &  1.7247 &  0.2840 &  0.1362 &  0.3628 \\
 26.85 &  9.4601 &  9.3653 &  7.0452 &  1.7319 &  0.2842 &  0.1344 &  0.3672 \\
 26.90 &  9.4546 &  9.3596 &  7.0507 &  1.7311 &  0.2847 &  0.1324 &  0.3686 \\
 26.95 &  9.4954 &  9.4001 &  7.0817 &  1.7384 &  0.2841 &  0.1347 &  0.3659 \\
 27.00 &  9.5138 &  9.4183 &  7.0968 &  1.7355 &  0.2838 &  0.1353 &  0.3677 \\
 27.05 &  9.5624 &  9.4665 &  7.1354 &  1.7193 &  0.2830 &  0.1361 &  0.3619 \\
 27.10 &  9.5786 &  9.4828 &  7.1538 &  1.7142 &  0.2825 &  0.1358 &  0.3607 \\
 27.15 &  9.5893 &  9.4933 &  7.1649 &  1.7236 &  0.2825 &  0.1359 &  0.3622 \\
 27.20 &  9.5758 &  9.4799 &  7.1710 &  1.7298 &  0.2829 &  0.1357 &  0.3665 \\
 27.25 &  9.5888 &  9.4927 &  7.1749 &  1.7309 &  0.2830 &  0.1355 &  0.3628 \\
 27.30 &  9.6558 &  9.5590 &  7.2252 &  1.7073 &  0.2817 &  0.1349 &  0.3530 \\
 27.35 &  9.6169 &  9.5203 &  7.2105 &  1.7218 &  0.2828 &  0.1353 &  0.3613 \\
 27.40 &  9.6383 &  9.5414 &  7.2330 &  1.7156 &  0.2823 &  0.1351 &  0.3599 \\
 27.45 &  9.6787 &  9.5819 &  7.2641 &  1.7229 &  0.2815 &  0.1338 &  0.3473 \\
 27.50 &  9.6752 &  9.5779 &  7.2594 &  1.7177 &  0.2818 &  0.1352 &  0.3584 \\
 27.55 &  9.7201 &  9.6225 &  7.3008 &  1.7114 &  0.2811 &  0.1341 &  0.3518 \\
 27.60 &  9.6626 &  9.5653 &  7.2764 &  1.7177 &  0.2823 &  0.1349 &  0.3605 \\
 27.65 &  9.6750 &  9.5781 &  7.2802 &  1.7145 &  0.2825 &  0.1350 &  0.3598 \\
 27.70 &  9.7340 &  9.6364 &  7.3311 &  1.6977 &  0.2810 &  0.1333 &  0.3473 \\
 27.75 &  9.7539 &  9.6559 &  7.3538 &  1.7018 &  0.2808 &  0.1332 &  0.3481 \\
 27.80 &  9.7842 &  9.6860 &  7.3670 &  1.7059 &  0.2805 &  0.1330 &  0.3449 \\
 27.85 &  9.7828 &  9.6845 &  7.3854 &  1.7005 &  0.2806 &  0.1326 &  0.3453 \\
 27.90 &  9.7891 &  9.6909 &  7.3967 &  1.6994 &  0.2806 &  0.1323 &  0.3433 \\
 27.95 &  9.9234 &  9.8237 &  7.4766 &  1.6746 &  0.2780 &  0.1287 &  0.3217 \\
\enddata
\end{deluxetable*}

\end{document}